\documentclass[twocolumn]{aastex631}
\usepackage[none]{hyphenat}
\usepackage{amsmath}  

\received{ }
\revised{ }
\accepted{}
\submitjournal{ApJ}

\shorttitle{Bars in clusters}
\shortauthors{Tawfeek, A. et al.}

\usepackage{graphicx}	
\usepackage{amsmath}	
\usepackage{amssymb}	
\usepackage[caption=false]{subfig}
\usepackage{float}
\usepackage{xcolor}

\begin{document}

\title{Morphology driven evolution of barred galaxies in OMEGAWINGS Clusters}

\correspondingauthor{Amira A. Tawfeek}
\email{amira.t.gabr@gmail.com}
\author[0000-0002-8279-9236]{Amira A. Tawfeek}
\affiliation{Instituto de Radioastronom\'ia y Astrof\'isica, Universidad Nacional Aut\'onoma de M\'exico, Antigua Carretera a P\'atzcuaro \# 8701, Ex-Hda. San Jos\'e de la Huerta, Morelia, Michoac\'an, M\'exico C.P. 58089}
\affiliation{National Research Institute of Astronomy and Geophysics (NRIAG), 11421 Helwan, Cairo, Egypt}

\author[0000-0002-2897-9121]{Bernardo Cervantes Sodi}
\affiliation{Instituto de Radioastronom\'ia y Astrof\'isica, Universidad Nacional Aut\'onoma de M\'exico, Antigua Carretera a P\'atzcuaro \# 8701, Ex-Hda. San Jos\'e de la Huerta, Morelia, Michoac\'an, M\'exico C.P. 58089}

\author[0000-0002-7042-1965]{Jacopo Fritz}
\affiliation{Instituto de Radioastronom\'ia y Astrof\'isica, Universidad Nacional Aut\'onoma de M\'exico, Antigua Carretera a P\'atzcuaro \# 8701, Ex-Hda. San Jos\'e de la Huerta, Morelia, Michoac\'an, M\'exico C.P. 58089}

\author[0000-0002-1688-482X]{Alessia Moretti}
\affiliation{INAF–Osservatorio astronomico di Padova, Vicolo dell’Osservatorio 5, 35122 Padova, Italy}

\author[0000-0002-4507-9571]{David P\'erez-Mill\'an}
\affiliation{Instituto de Radioastronom\'ia y Astrof\'isica, Universidad Nacional Aut\'onoma de M\'exico, Antigua Carretera a P\'atzcuaro \# 8701, Ex-Hda. San Jos\'e de la Huerta, Morelia, Michoac\'an, M\'exico C.P. 58089}

\author[0000-0002-7296-9780]{Marco Gullieuszik}
\affiliation{INAF–Osservatorio astronomico di Padova, Vicolo dell’Osservatorio 5, 35122 Padova, Italy}

\author[0000-0001-8751-8360]{Bianca M. Poggianti}
\affiliation{INAF–Osservatorio astronomico di Padova, Vicolo dell’Osservatorio 5, 35122 Padova, Italy}

\author[0000-0003-0980-1499]{Benedetta Vulcani}
\affiliation{INAF–Osservatorio astronomico di Padova, Vicolo dell’Osservatorio 5, 35122 Padova, Italy}

\author[0000-0002-4158-6496]{Daniela Bettoni}
\affiliation{INAF–Osservatorio astronomico di Padova, Vicolo dell’Osservatorio 5, 35122 Padova, Italy}


\begin{abstract}
We present a study of barred galaxies in the cluster environment, exploiting a sample of galaxies drawn from the extended WIde-field Nearby Galaxy-cluster Survey (OmegaWINGS) that covers up to the outer regions of 32 local X-ray selected clusters. Barred galaxies are identified through a semi-automatic analysis of ellipticity and position angle profiles. We find, in agreement with previous studies, a strong co-dependence of the bar fraction with the galaxy stellar mass and morphological type, being maximum for massive late-type galaxies. The fraction of barred galaxies decreases with increasing cluster mass and with decreasing clustercentric distance, a dependence that vanishes once we control for morphological type, which indicates that the likelihood of a galaxy hosting a bar in the cluster environment is determined by its morphological transformation.
At large clustercentric distances, we detect a dependence on the distance to the nearest neighbor galaxy, suggesting that tidal forces with close companions are able to suppress the formation of bars or even destroy them. Barred galaxies in our sample are either early-type, star forming galaxies located within the virial radii of the clusters or late-type quenched galaxies found beyond the virial radii of the clusters. We propose a scenario in which already quenched barred galaxies that fall into the clusters are centrally rejuvenated by the interplay of the perturbed gas by ram-pressure and the bar, in galaxies that are undergoing a morphological transformation.
\end{abstract}

\section{Introduction} \label{sec:intro}
Stellar bars are a common structure found in a large fraction of disk galaxies (early and late type) \citep{Nair2010, Masters2011, Fraser2020}. They are formed mainly when the stellar orbits of disk galaxies deviate from a circular path due to instabilities \citep{Athanassoula1983, Contopoulos1989, Pfenning1991}. Detecting such structure is important in tracing the formation and the evolution of galaxies where bars are able to transport a massive amount of gas (i.e materials) between the disk and the bulge of the host galaxy \citep{Ho1997, Combes2003, Kormendy2004, Kim2020}.

Bars are present nearly in 60$\%$ of disk galaxies in the local universe \citep{Eskridge2000, Marinova2007, Buta2010, Ann2015}. They are formed in different environments such as isolated galaxies, pairs, triplets, groups and clusters of galaxies \citep{Jogee2004, Sheth2008}.

Hydrodynamic simulations indicate that bars have a clear role in driving the evolution of host galaxies as they are considered to be an important channel through which gas can be transferred from the outer parts of the disk into the central region of the host galaxy. This provides a quite efficient mechanism to also fuel star formation, building central bulges, feeding massive black holes and hence favoring AGN activity, and drive interaction between stellar disk and dark matter halo \citep{Athanassoula2003, Dalcanton2004, Sellwood2014}. 

Several studies have shown that barred galaxies are characterized by a higher central star formation (SF) activity than unbarred galaxies \citep{Jogee2005, Hunt2008, Ellison2011, Wang2012}. On the other side, \citet{James2015, James2018, George2019,George2020} have found that stellar bars can quench the SF activity due to the redistribution of cold gas in the bar region.

Bars are also thought to have a great impact on the formation of bulges, on the redistribution of angular momentum and mass of disk galaxies \citep{Friedli1995, Athanassoula2002, Sheth2003, Aguerri2009,Sellwood2014}, and the accelerated formation of spiral structures \citep{Lynden1972, Pfenning1991}. Hence, understanding the prevalence of bars as a function of galaxy properties and the surrounding environment is important in tracing the formation and the evolution of galaxies \citep{Ho1997, Combes2003, Kormendy2004, Kim2017, Spinoso17, Kim2020}.

Galaxy clusters are known to be very efficient agents in producing deep changes on the properties of galaxies. Once galaxies enter such dense regions, both their stellar populations and morphology are severely affected: they rapidly turn into passive, early type objects as a result of the variety of physical processes acting on them. The changes in morphology are particularly evident, by various tokens such as the so-called ``morphology-density relation'' \citep[e.g.][]{dressler80, fasano15} and the Butcher-Oemler effect \citep{butcher84, poggianti09}. 

In previous studies of the Coma cluster, \citet{Marinova2012} and \citet{Lansbury2014} found a very weak correlation between clustercentric distance and bar fraction, with a low significance increase of bars towards the cluster core. \cite{Mendez-Abreu2012}, studying galaxies in three different environment, raging from field to galaxies belonging to the Coma and Virgo clusters, concluded that for bright galaxies bar formation and evolution depend mostly on internal processes rather than on environmental ones. On a comprehensive study using different statistics to account for the local and large scale environment, \cite{Lin2014} found indications that interactions with close companions can effectively suppress the formation and growth of stellar bars, as previously reported by \cite{Lee2012}, and no evidence of environmental stimulation for their formation, in contrast with early studies \citep{Elmegreen1990}. Similarly, \cite{Sarkar2021}, using an information theoretical framework to look for correlations between the presence of bars and their environment, concluded that the formation of a bar is mostly determined by internal processes, independent of their environment. In contrast, \citep{Lokas2016}, using N-body simulations of galaxies orbiting a Virgo-like cluster, indicated that the cluster tidal force can trigger bar formation in the cluster core but not in the outskirts, and hence a larger concentration of barred galaxies can be found in the cluster center, a work later expanded in \cite{Lokas2020}.

In this study we aim to identify the stellar bars in galaxies belonging to 32 local clusters drawn from the OmegaWINGS survey and study their dependence on the intrinsic galaxy properties (the stellar mass, the (B-V) color and the morphological type) as well as the global properties of the galaxy clusters (the X-ray luminosity) and the cluster environment (the projected clustercentric distance ($(r_p/r_{virial})_{cl}$) and the projected distance to the nearest neighbor galaxy($r_p/r_{virial}$)$_{ng}$). Finally, we study how the presence of a bar affects the star formation activity in cluster galaxies.

This paper is organized as follows: in section~\ref{sec:sample} we introduce the main dataset of galaxy clusters drawn from OmegaWINGS. Section~\ref{sec:methodology} describes our strategy for bar detection and the basic requirements. The main results including the bar fraction and its dependence on the properties of the host galaxies as well as the dependence of bar fraction on the cluster environment are presented in Section~\ref{sec:results}. Finally, our conclusions are given in Section~\ref{sec:conclusions}.

As for the cosmological parameters, we have adopted those commonly used in the frame of the WINGS and OmegaWINGS projects, following the $\Lambda$ cold dark matter model ($\Lambda$CDM) with $H_0= 70$ km s$^{-1}$ Mpc$^{-1}$, $\Omega_{\Lambda}$=0.7 and $\Omega_{M}=0.3$. The adopted IMF is the one of \cite{chabrier03} with stellar masses in the $0.1 - 100$ M$_\odot$ range.

\section{Observational data}
\label{sec:sample}

\subsection{The OmegaWINGS galaxy sample}

OmegaWINGS \citep{Gullieuszik2015} is an extension of the WIde-field Nearby Galaxy-cluster Survey \citep[WINGS,][]{Fasano2006} multi-wavelength survey that mapped local galaxy clusters within the redshift range 0.04$<$z$<$0.07. The WINGS survey homogeneously covers, within one single image, a 34$\times$ 34 arcmin$^2$ field of view (FOV) with roughly 1.5 magnitude deeper than other sky survey such as the Sloan Digital Sky Survey (SDSS). The original WINGS survey was based on B- and V- imaging for the 76 clusters taken with the Wide Field Cameras on the INT (Isaac Newton Telescope) and the 2.2m MPG/ESO telescopes \citep{Varela2009}. This FOV covers only the core of the clusters and is hence missing the external regions at distances larger than the clusters virial radius. For that reason the WINGS survey has been extended to OmegaWINGS survey.

OmegaWINGS optical images are based on OmegaCAM B- and V- observations with the VLT (Very Large Telescope) Survey Telescope \citep[VST,][]{Capaccioli2011} in which one square degree covers not just the core but also the outer regions of the clusters, going beyond the cluster virial radius \citep{Gullieuszik2015}. 

OmegaWINGS clusters were selected from the 57 WINGS clusters that can be observed from the VST in B- and V- band imaging with OmegaCAM, a camera that samples the 1 deg$^2$ VST unvignetted FOV with a mosaic of 32 4k$\times$2k CCDs at 0$\arcsec$.21$/$pix. Under these conditions the photometric observation of 46 OmegaWINGS clusters were covered successfully \citep{Gullieuszik2015}. 
The spectroscopic observations of the 46 OmegaWINGS clusters were carried by using the AAOmega spectrograph \citep{Smith2004, Sharp2006} at the Australian Astronomical Observatory (AAT), which can host up to 392 fibers over 2 $\times$ 2 deg$^2$ with a diameter of 2.16 $\arcsec$ for each fiber \citep[see][for more details]{Moretti2017}.

OmegaWINGS clusters span almost two orders of magnitude in X-ray luminosity (L$_X \sim 0.2 - 5 \times 10^{44}$ ergs s$^{-1}$), that roughly correspond to a dynamical mass of $\sim 10^{14}$ to more than $10^{15} M_\odot$ following \cite{Reiprich02}. The cluster size is characterized by the virial radius, defined as the radius where the mean overdensity drops to 200 times the critical density of the Universe, computed using the velocity dispersion of the cluster as provided by \cite{Cava2009} and the formula \citep{Finn2005}:

\begin{equation}
    r_{\textrm{virial}}=1.75 \times \frac{\sigma}{1000 \textrm{ km} \textrm{ s}^{-1}} \frac{1}{\sqrt{\Omega_\Lambda + \Omega_0(1+z)^3 }} h^{-1} \textrm{(Mpc)}.
\end{equation}

\subsection{Physical properties of galaxies} \label{sec:properties}
In this work, we will use stellar population properties, such as the total mass, the star formation history, and optical rest-frame colors, as calculated by Pérez-Millán et al. (submitted) exploiting the spectrophotometric fitting code {\sc sinopsis}, whose characteristics are described in details in \cite{fritz07,fritz17}. The adopted IMF is the one of \cite{chabrier03} with stellar masses in the $0.1 - 100$ M$_\odot$ range; nebular gas emission is added to the theoretical spectra, and is modeled as described in \cite{fritz17}, thus allowing the calculation of the recent star formation rate (SFR) within {\sc sinopsis} by fitting observed emission lines. Such spectral analysis also allows to reliably recover the star formation history (SFH, the star formation rate as a function of the cosmic time) in four age bins:
\begin{enumerate}
    \item $0-2\times 10^{7}$ yr
    \item $2\times 10^{7}-5.6\times 10^{8}$ yr
    \item $5.6\times 10^{8}-5.6\times 10^{9}$ yr
    \item $5.6\times 10^{9}-T_{gal}$ yr
\end{enumerate}
where $T_{gal}$ is the age of the oldest stellar populations, limited by the age of the universe at the galaxy's redshift. Note that the first age bin is the one in which nebular emission lines due to star formation are observed, and is hence the one that is usually referred to as the present day (or even instantaneous) SFR.

It should be noted that the aforementioned properties are obtained from fiber spectra, hence sampling a limited region at the center of the galaxies, typically the 2-3 innermost kpc. While \cite{fritz11} have demonstrated that, with appropriate aperture and color corrections, the total stellar mass is properly recovered, the total SFR is a much more complicate issue, due to the intrinsically patchy nature in the geometrical distribution of star forming regions. Hence, color corrections might not be sufficient to properly recovery this particular property on the whole galactic scale. Nevertheless, the central regions are exactly those mostly affected by the presence of a bar, hence the derived properties of the stellar populations are mostly representative of the region of influence of the bar.

The morphological classification has been carried out on 39923 galaxies in 76 clusters of the WINGS survey by \cite{Fasano2012} with {\sc morphot}, an automated, non-parametric tool specifically designed to enhance the ability to distinguish between elliptical and S0 galaxies from the galaxy image by using neural network (NN) technique. In this study we segregate early-type as galaxies with -5 $ < M_{type} \le $ 0 and late-type galaxies with $M_{type} >$ 0.

\section{Sample selection and bar identification}
\label{sec:methodology}

The sample of clusters used in this study is based on the V-band images of 32 clusters drown from OmegaWINGS photometric catalog \citep{Gullieuszik2015} with spectroscopic information extracted from \citet{Moretti2017}. Statistical corrections are applied to account for both magnitude and geometrical completeness are calculated as in \cite{Cava2009}, and each galaxy will be weighted with the inverse of its corresponding completeness value. 

In order to exclude most elliptical galaxies and to avoid projection effect, we restricted our sample  to galaxies with morphological type M$_{type} >$ -5 and ellipticity $<$ 0.5. In addition, we rejected galaxies fainter than 21.5 mag/arcsec$^2$ in V-band, where isophotes are hard to be fitted in the outer most faint regions, and a possible bar detection would hence be uncertain.
Adapting this criteria we used for bar detection a sample of 32 OmegaWINGS clusters, with a total number of 3,456 galaxies, was identified.

Bars can be detected via various methods among which the most used ones are visual inspection, multi-component decomposition, Fourier analysis and ellipse fitting. Visual inspection is a very commonly adopted method that provides a quite high success rate of bar detection\citep{Sheth2008, Aguerri2009a, Buta2010}. However, this method is a time consuming and it is hence far from being the ideal one when dealing with a large sample. Multi-component decomposition, on the other hand, is based on an empirical technique and it is difficult to use with a large sample of galaxies spanning a large range of morphological types \citep{Lansbury2014}. Although Fourier analysis can be used as an automated method for bars identification, it is not able to classify weak bars, possibly leading to an underestimation of the barred galaxies fraction \citep{Lee2019}.

One widely used technique with a detection rate similarly to that of the visual inspection, is isophotal fitting through ellipses. From an observational point of view, bars affect the surface brightness properties of their hosting galaxies by producing isophotes that display a relatively constant (within $\pm$ 20$^\circ$) position angle (PA), and an ellipticity ($\epsilon$) monotonically increasing as a function of the galactocentric distance. At the end of the bar, the ellipticity drops sharply, and the PA changes as the isophotes belonging to the under-lying disk are fitted \citep{Wozniak1995,Martini2001,Laine2002, Sheth2003,Jogee2004}. 

Based on this technique, a galaxy can be classified as barred if a significant rise in the ellipticity ($\epsilon_{max}$ ) followed by a significant decrease ($\epsilon_{min}$ ) with variation $\Delta\epsilon$ $\geq 0.1$ is detected \citep{Wozniak1995, Martini2001,Laine2002, Sheth2003,Jogee2004}. 

\begin{figure*}
\centering
\begin{tabular}{lll}
\includegraphics[width=0.2\textwidth]{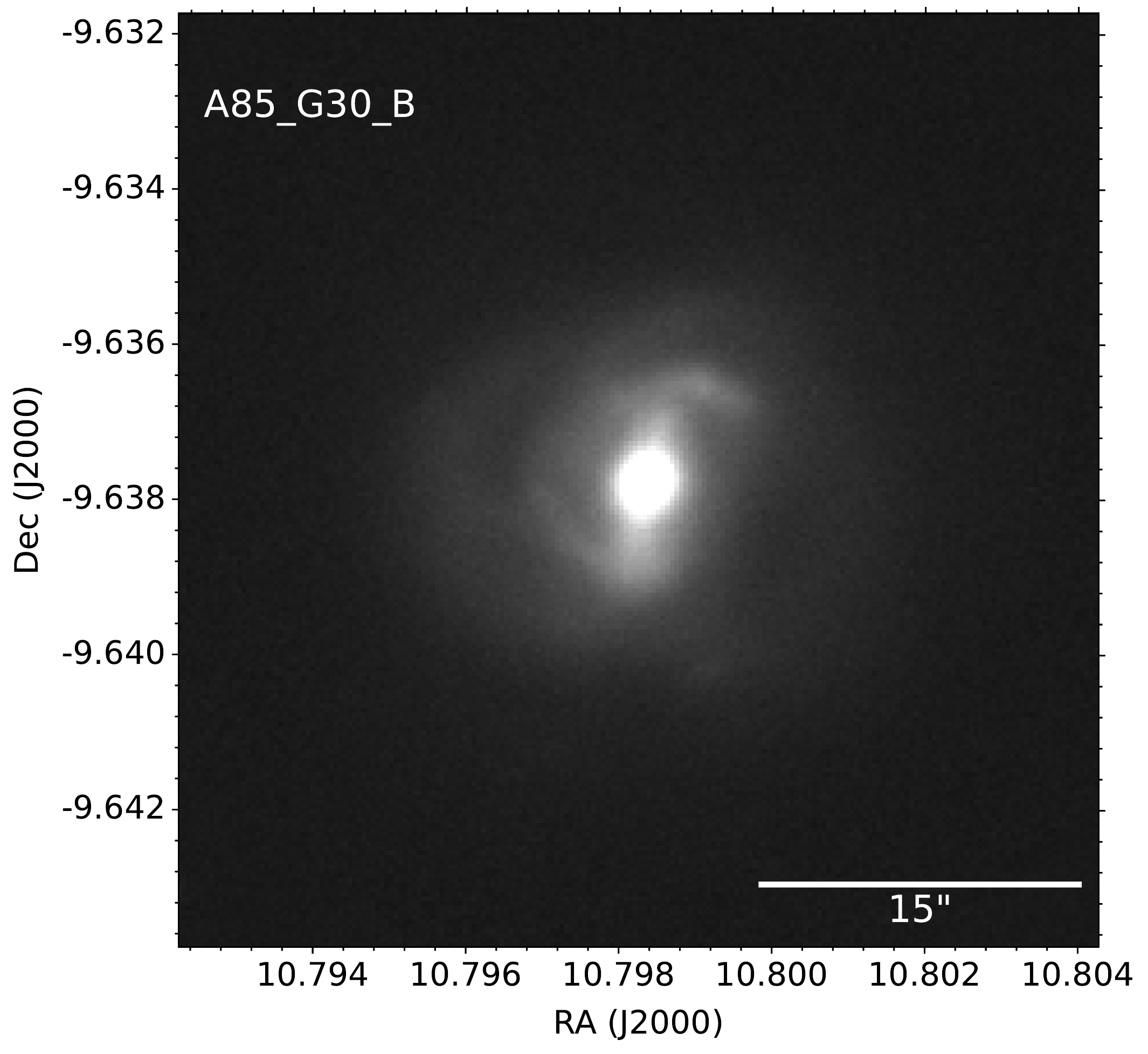} &
\includegraphics[width=0.2\textwidth]{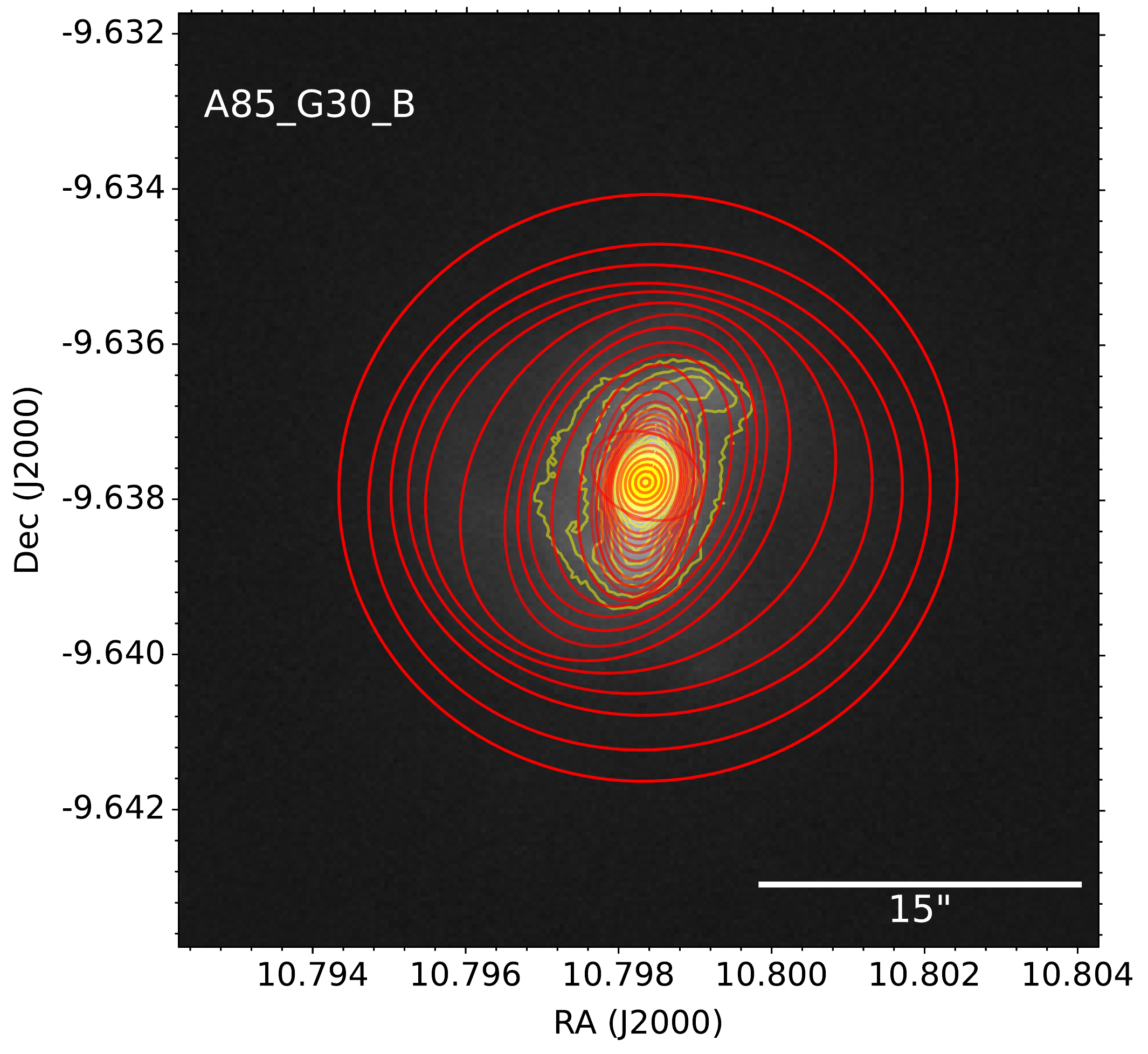} &
\includegraphics[width=0.28\textwidth]{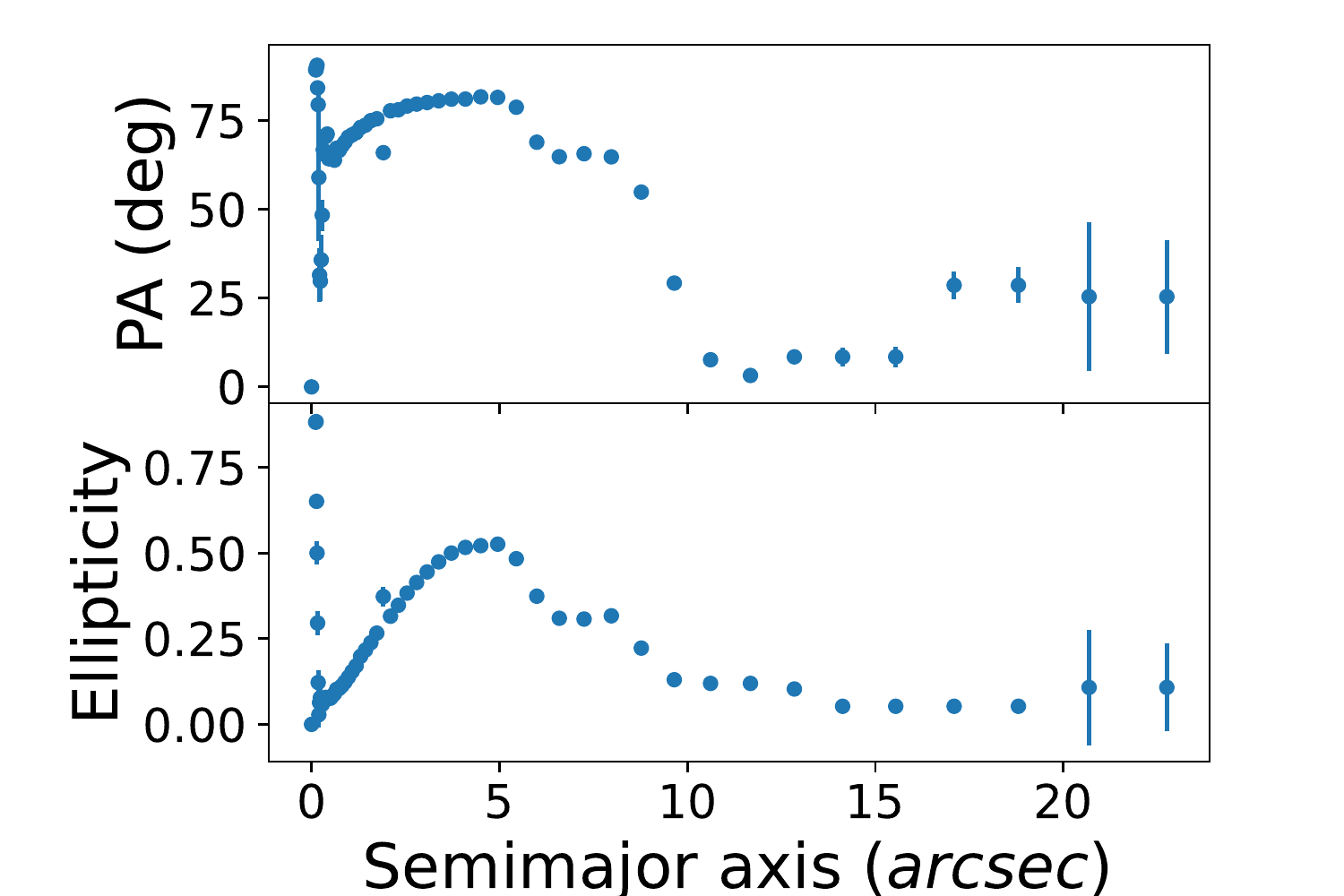} \\
\includegraphics[width=0.2\textwidth]{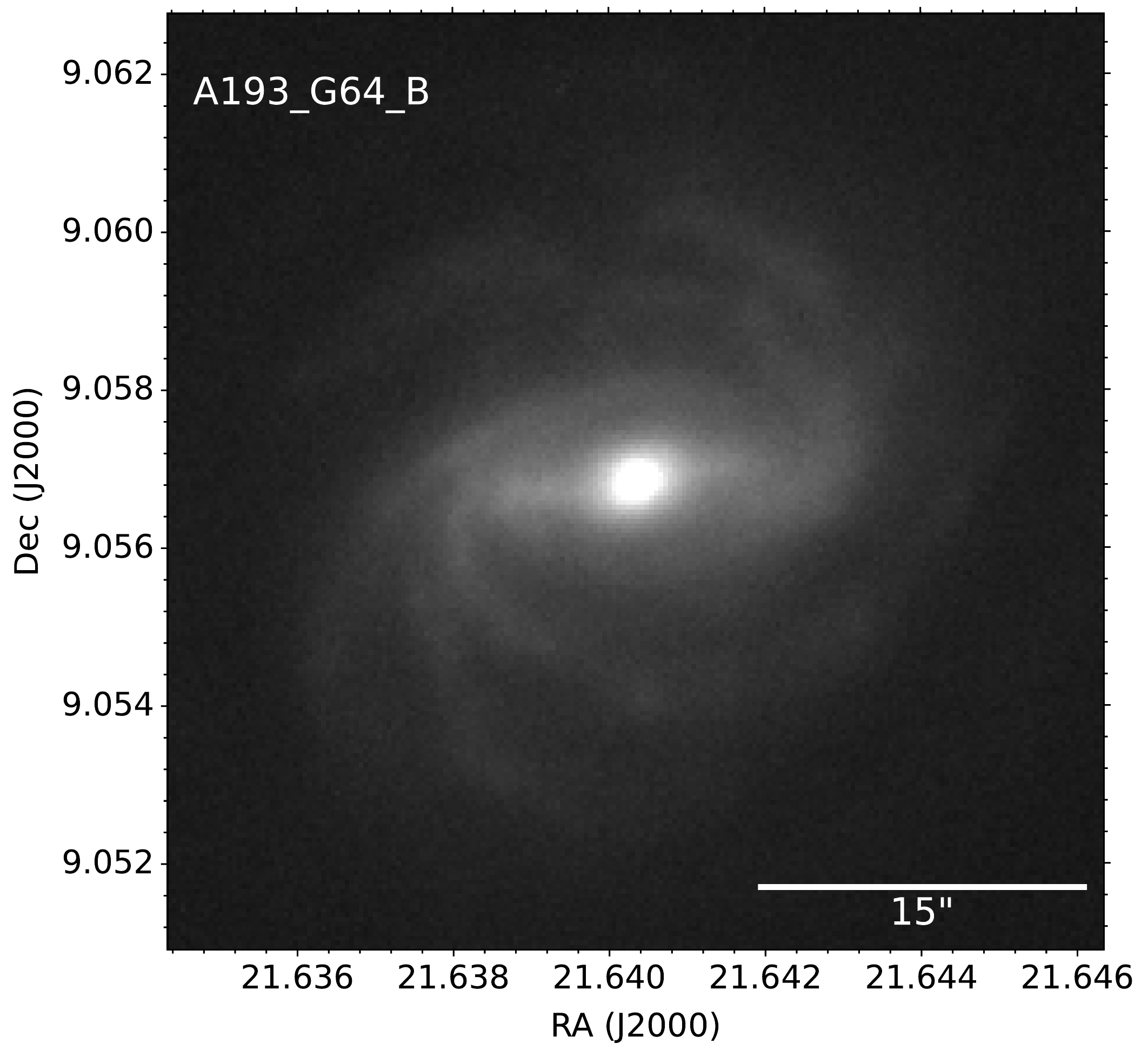} &
\includegraphics[width=0.2\textwidth]{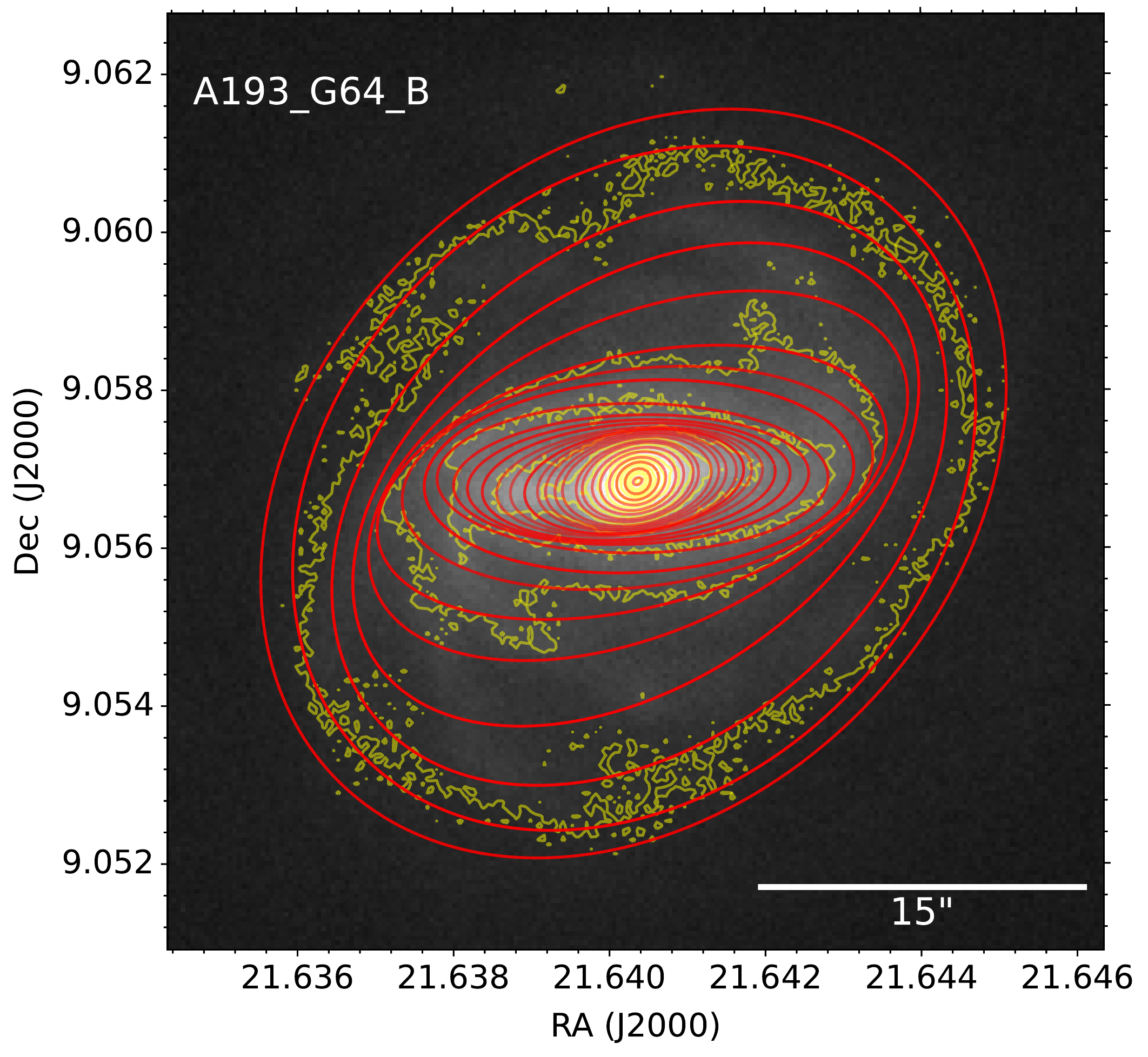} &
\includegraphics[width=0.28\textwidth]{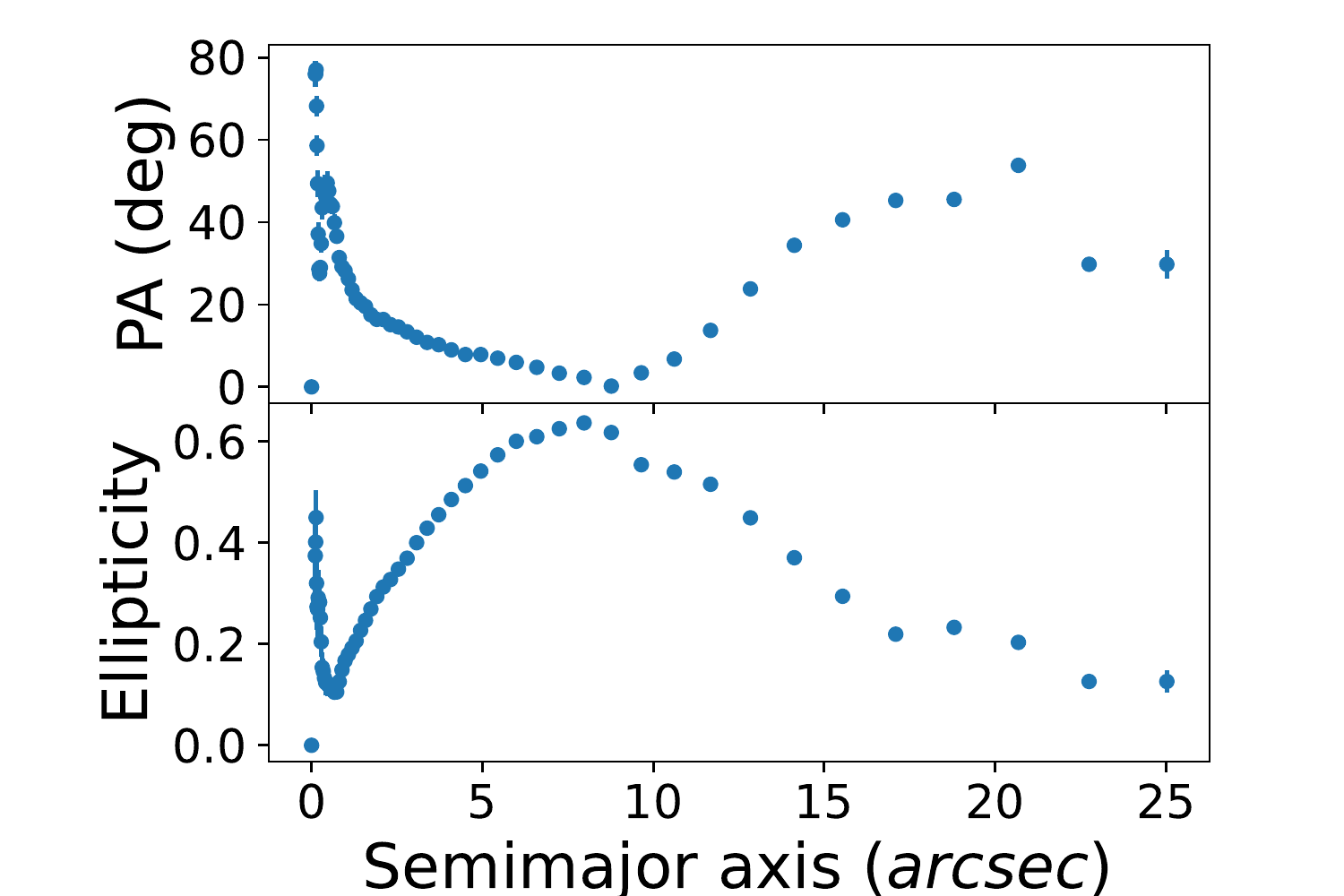} \\
\includegraphics[width=0.2\textwidth]{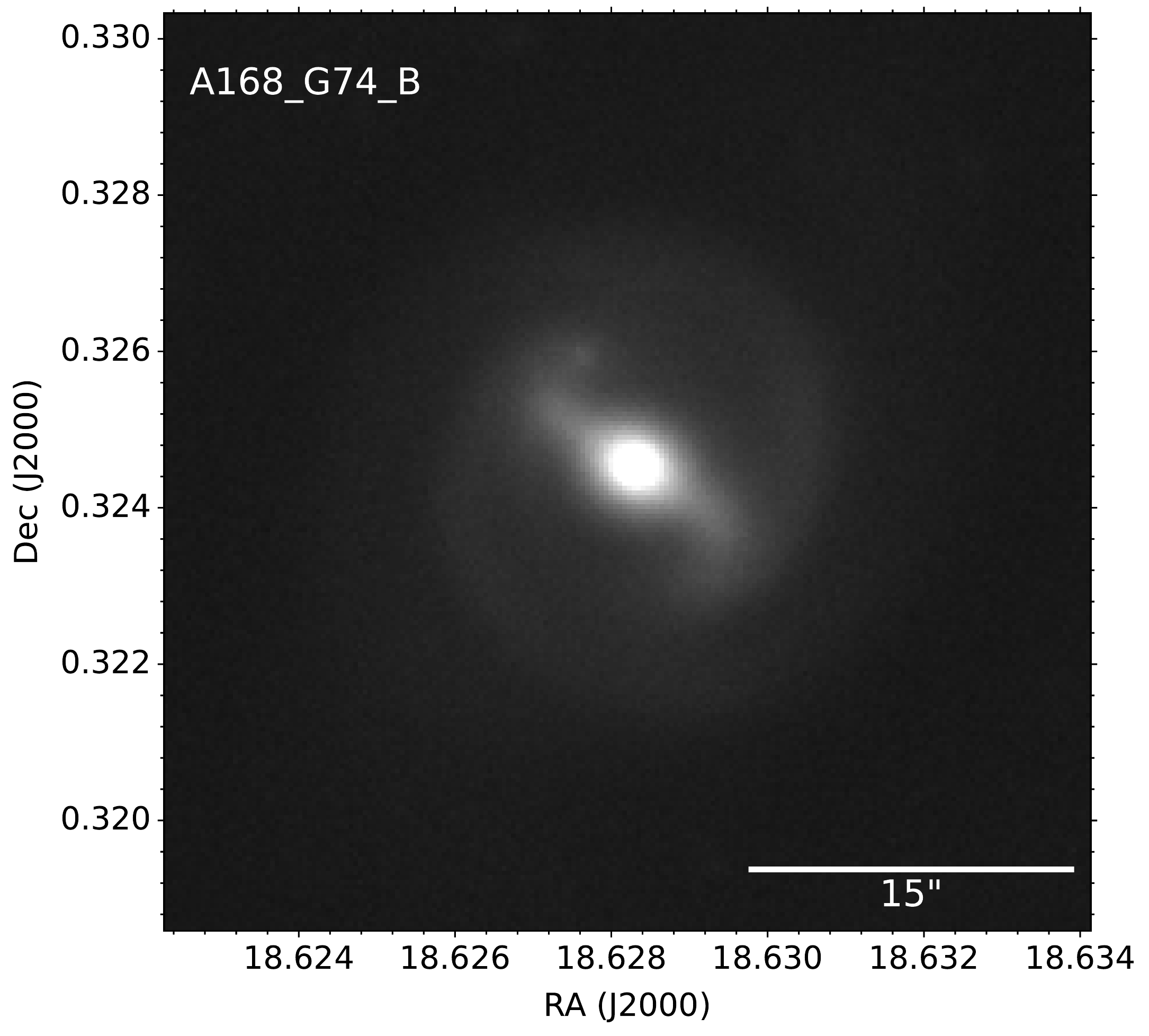} &
\includegraphics[width=0.2\textwidth]{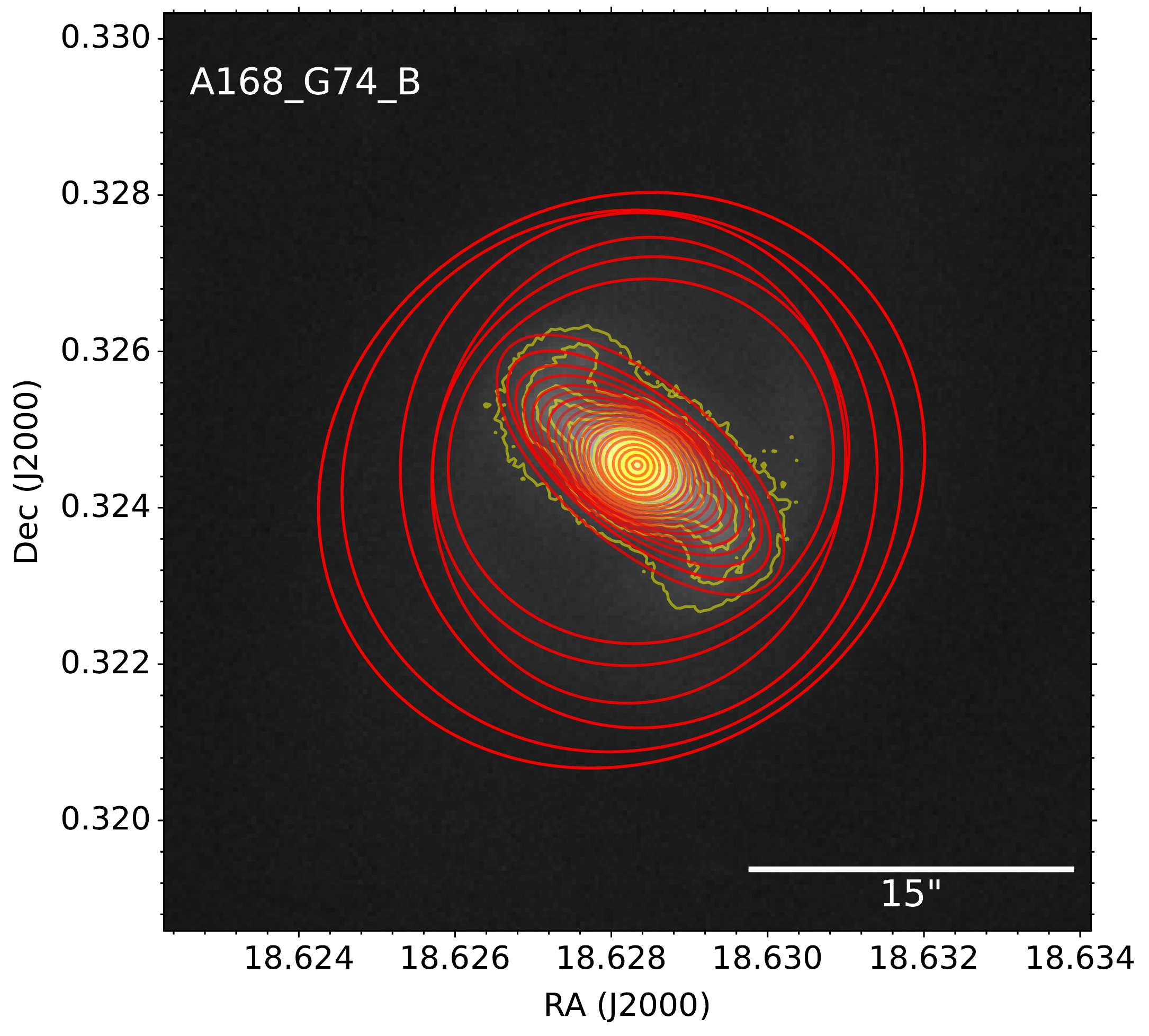} &
\includegraphics[width=0.28\textwidth]{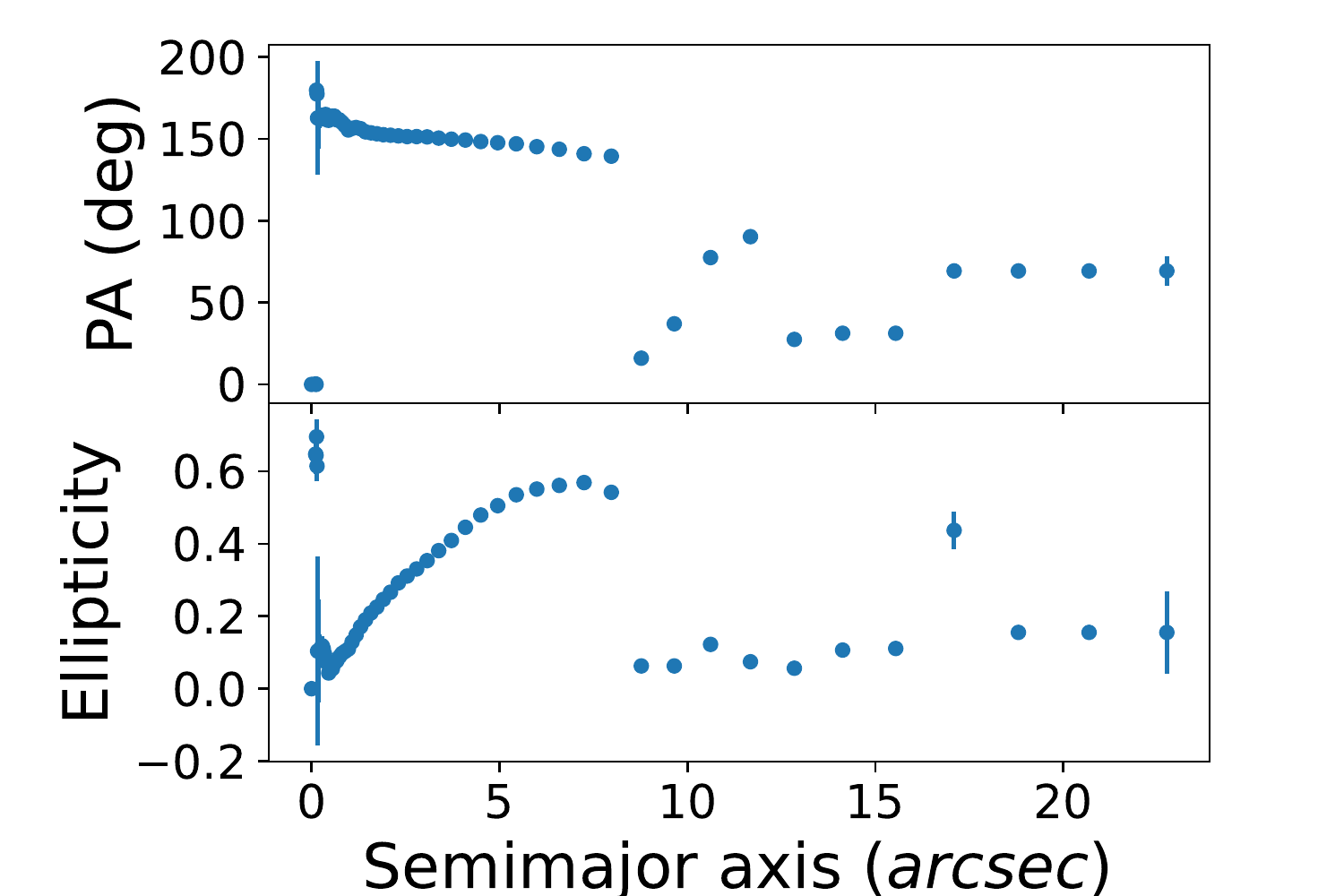} \\
\includegraphics[width=0.2\textwidth]{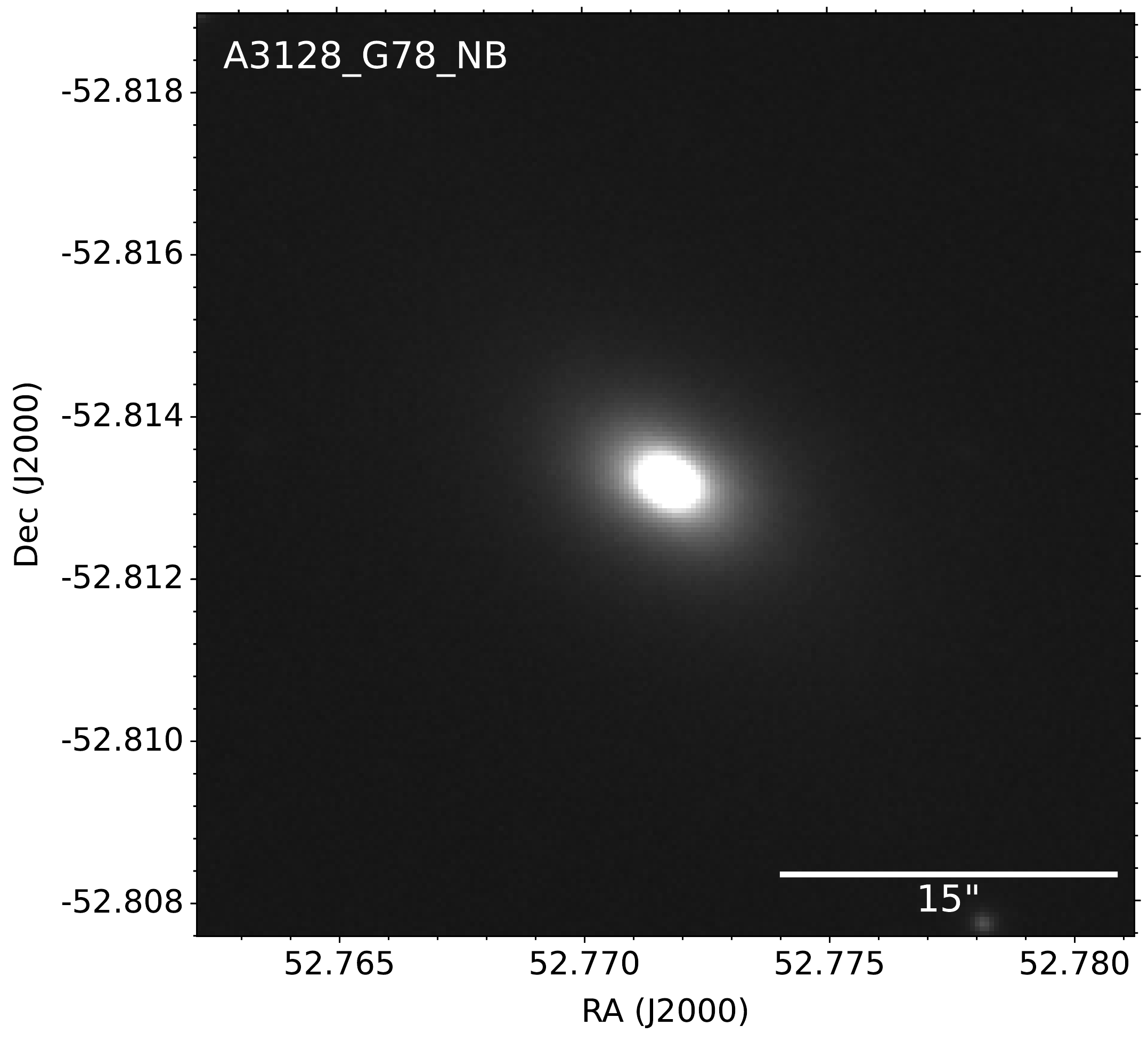} &
\includegraphics[width=0.2\textwidth]{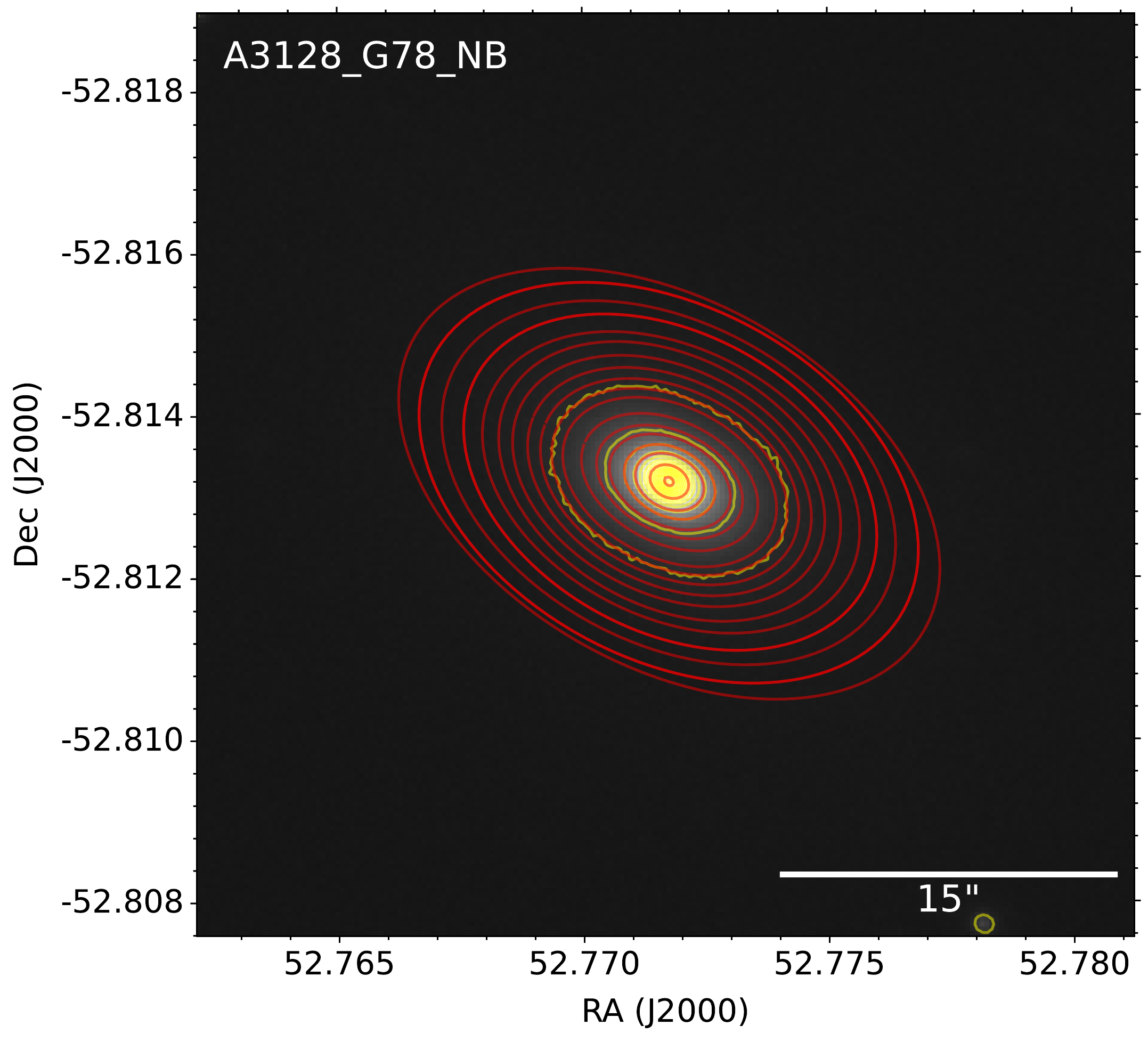} &
\includegraphics[width=0.28\textwidth]{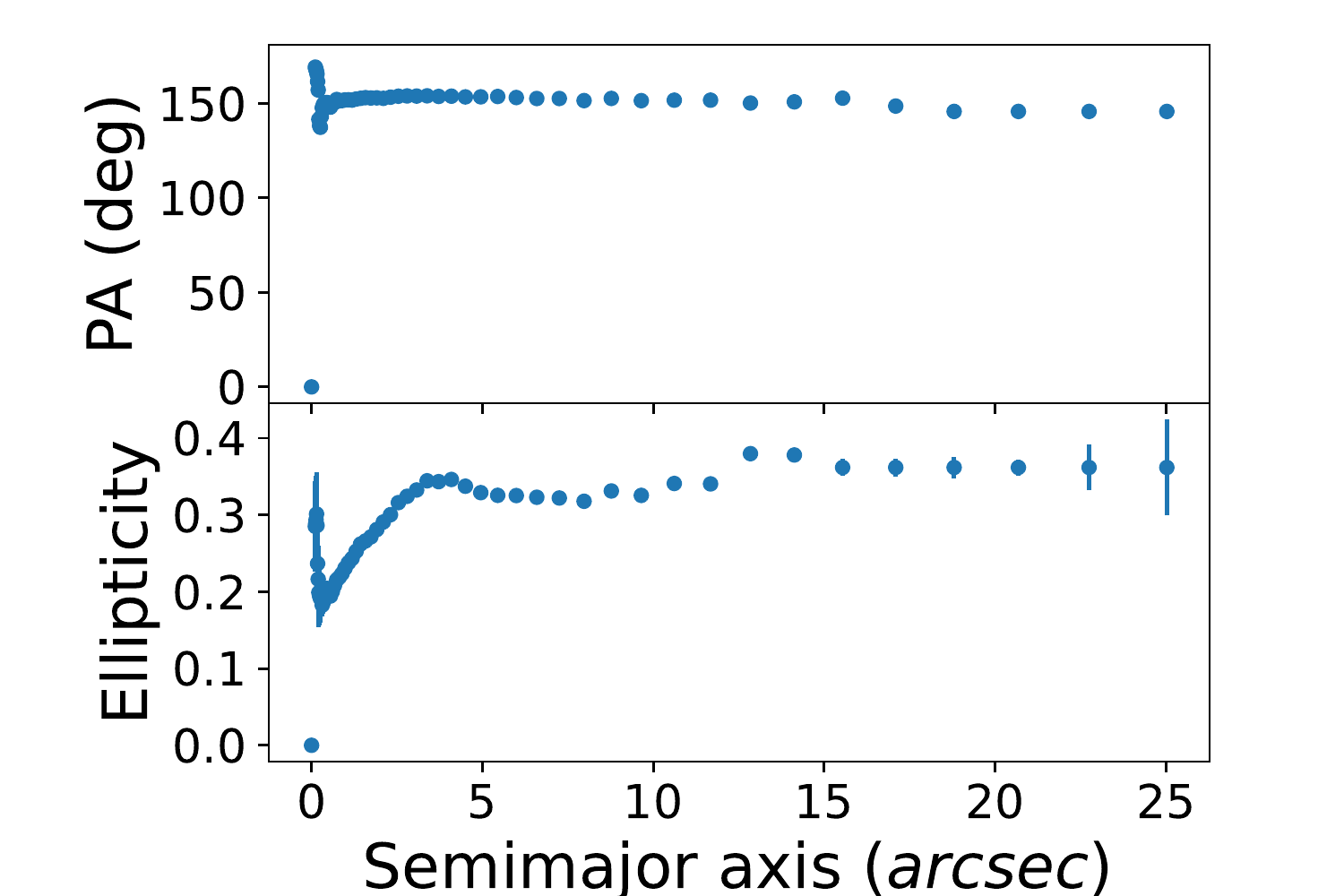} \\
\includegraphics[width=0.2\textwidth]{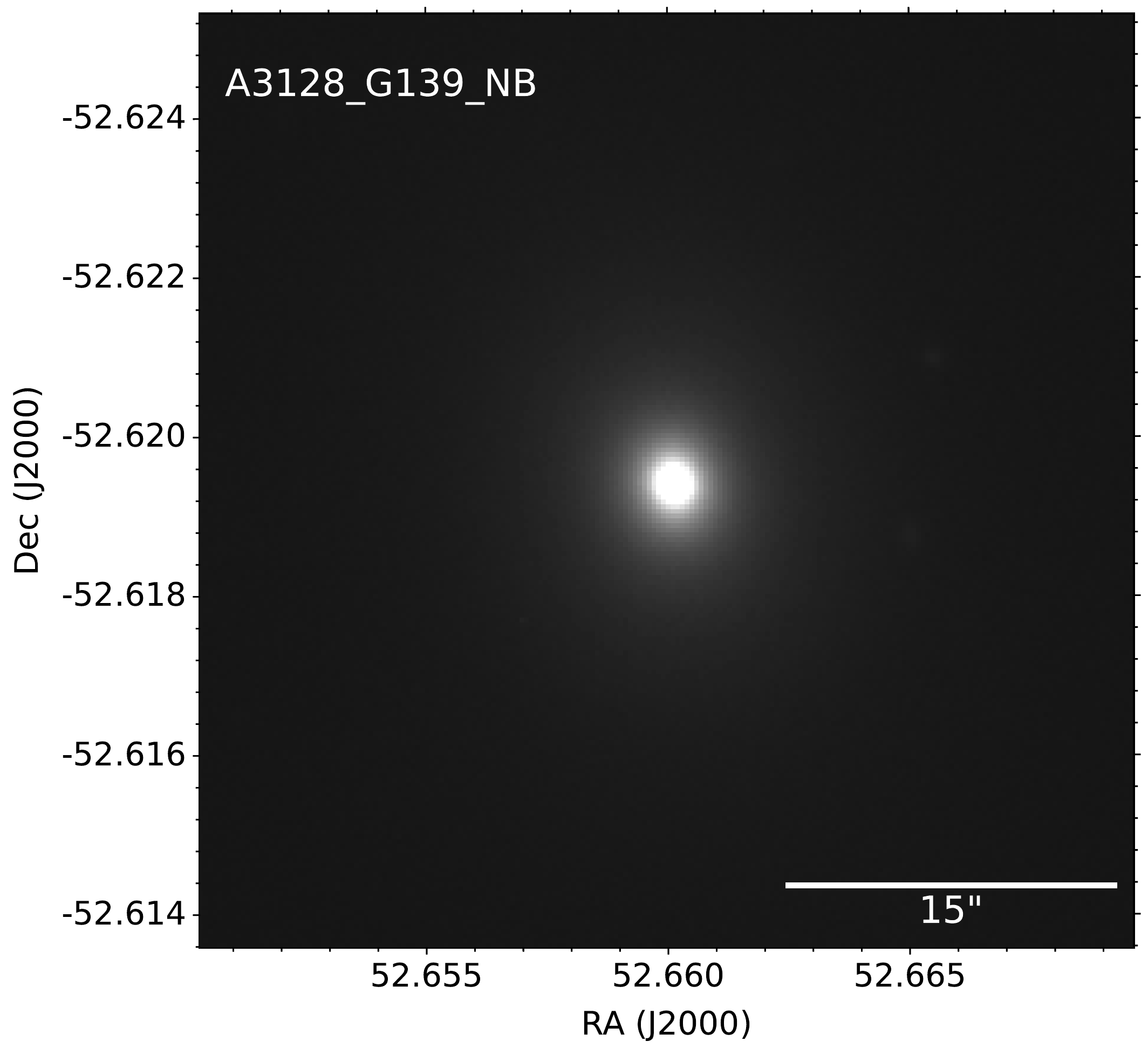} &
\includegraphics[width=0.2\textwidth]{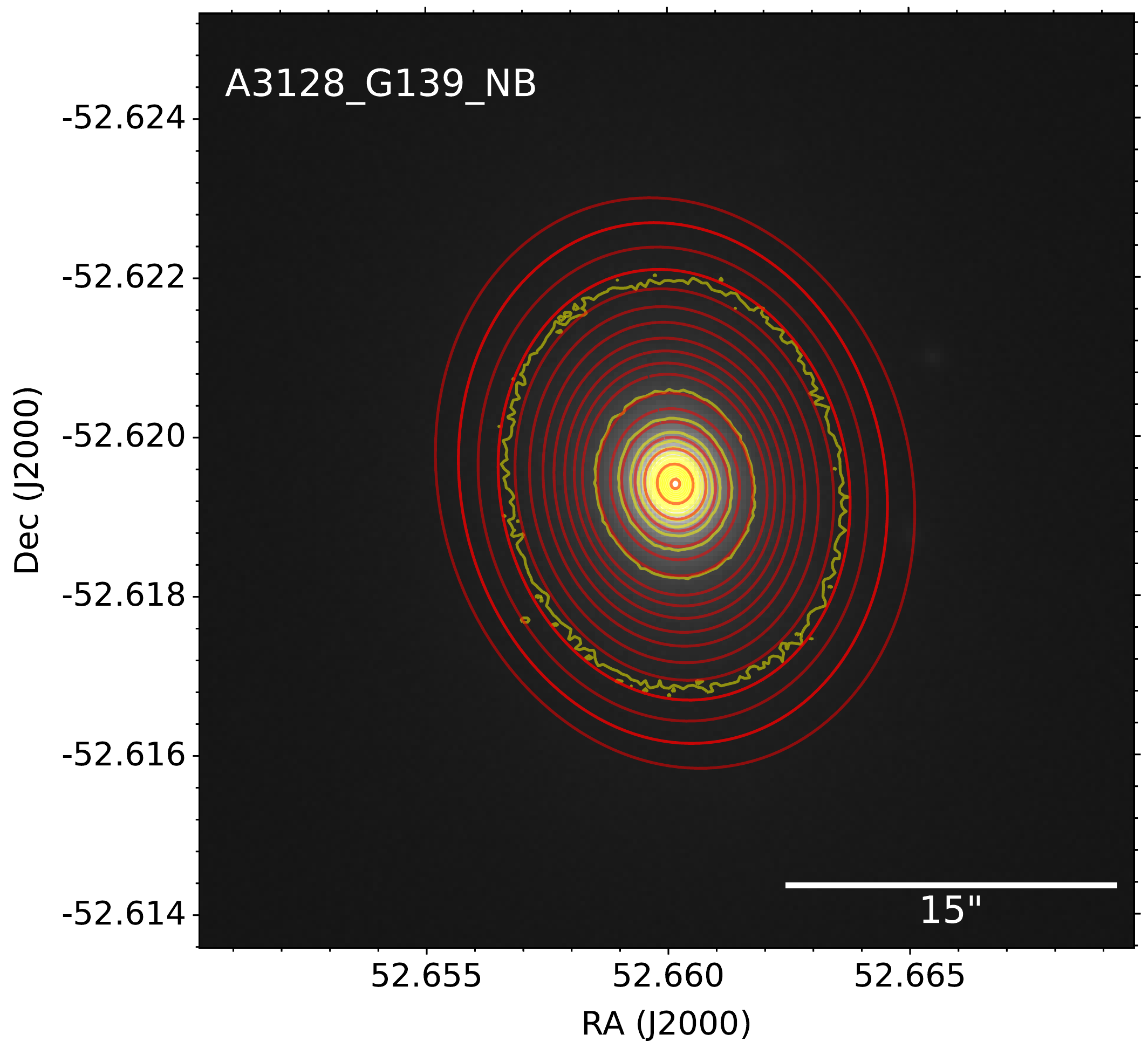} &
\includegraphics[width=0.28\textwidth]{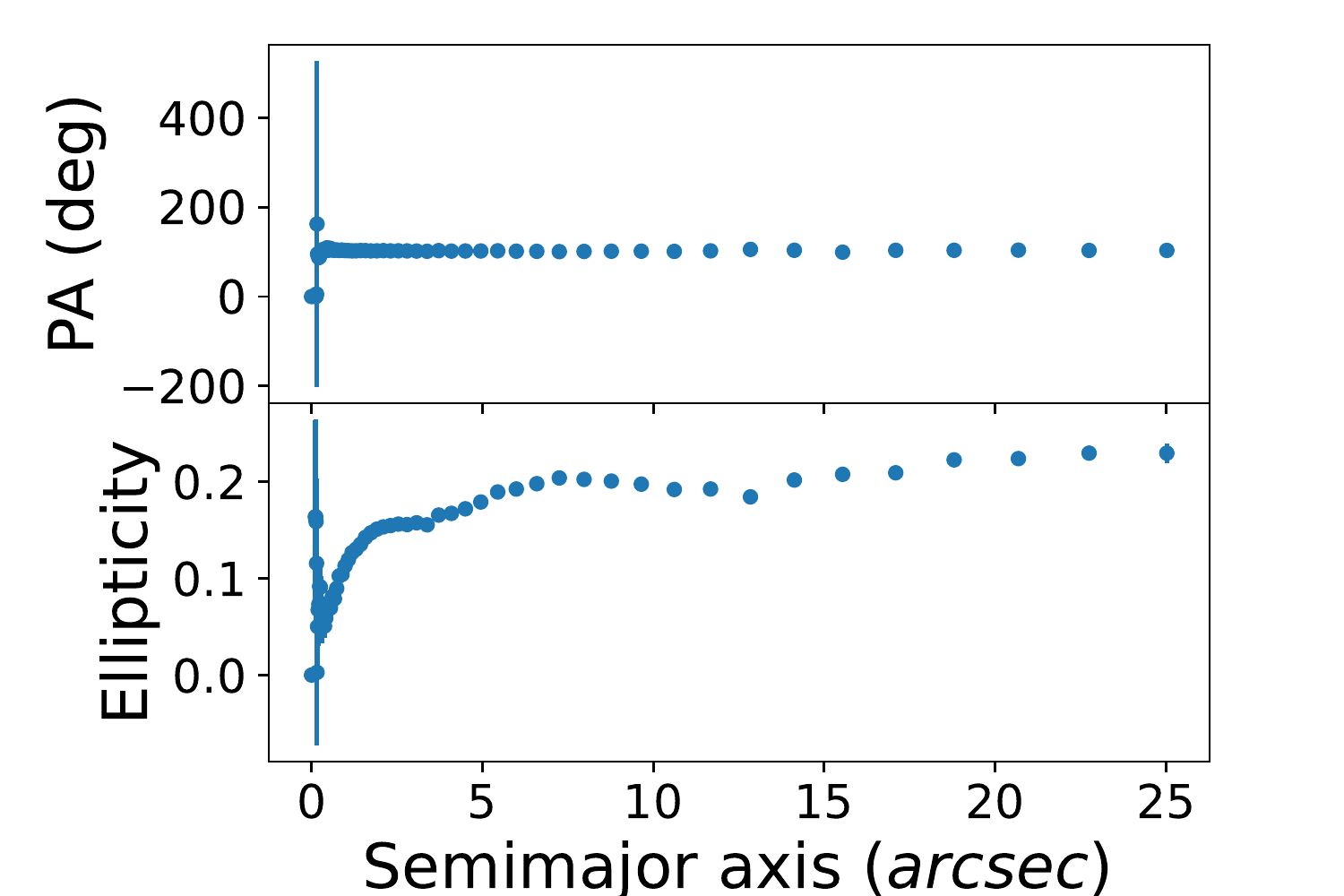} \\
\includegraphics[width=0.2\textwidth]{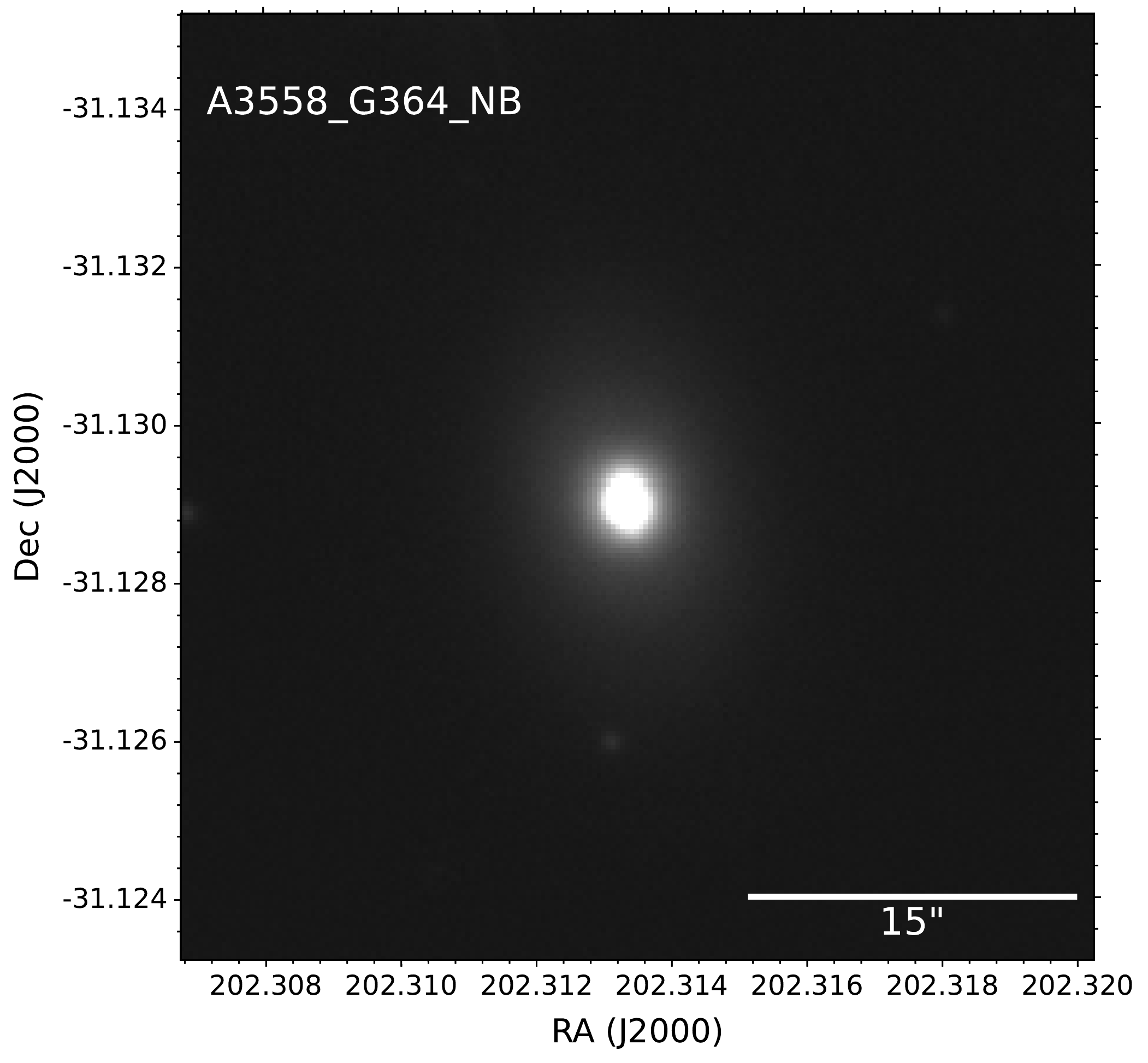} &
\includegraphics[width=0.2\textwidth]{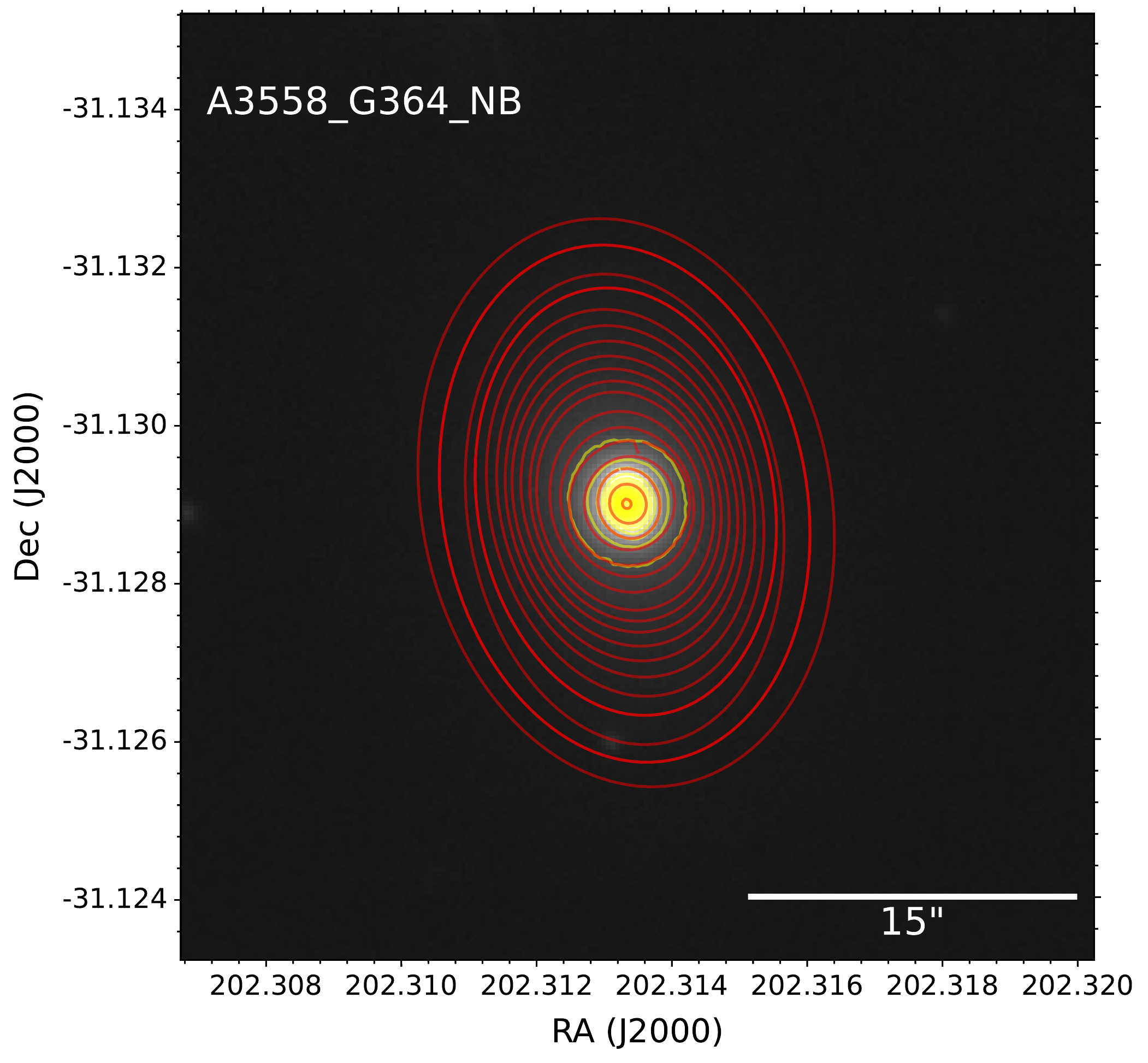} &
\includegraphics[width=0.28\textwidth]{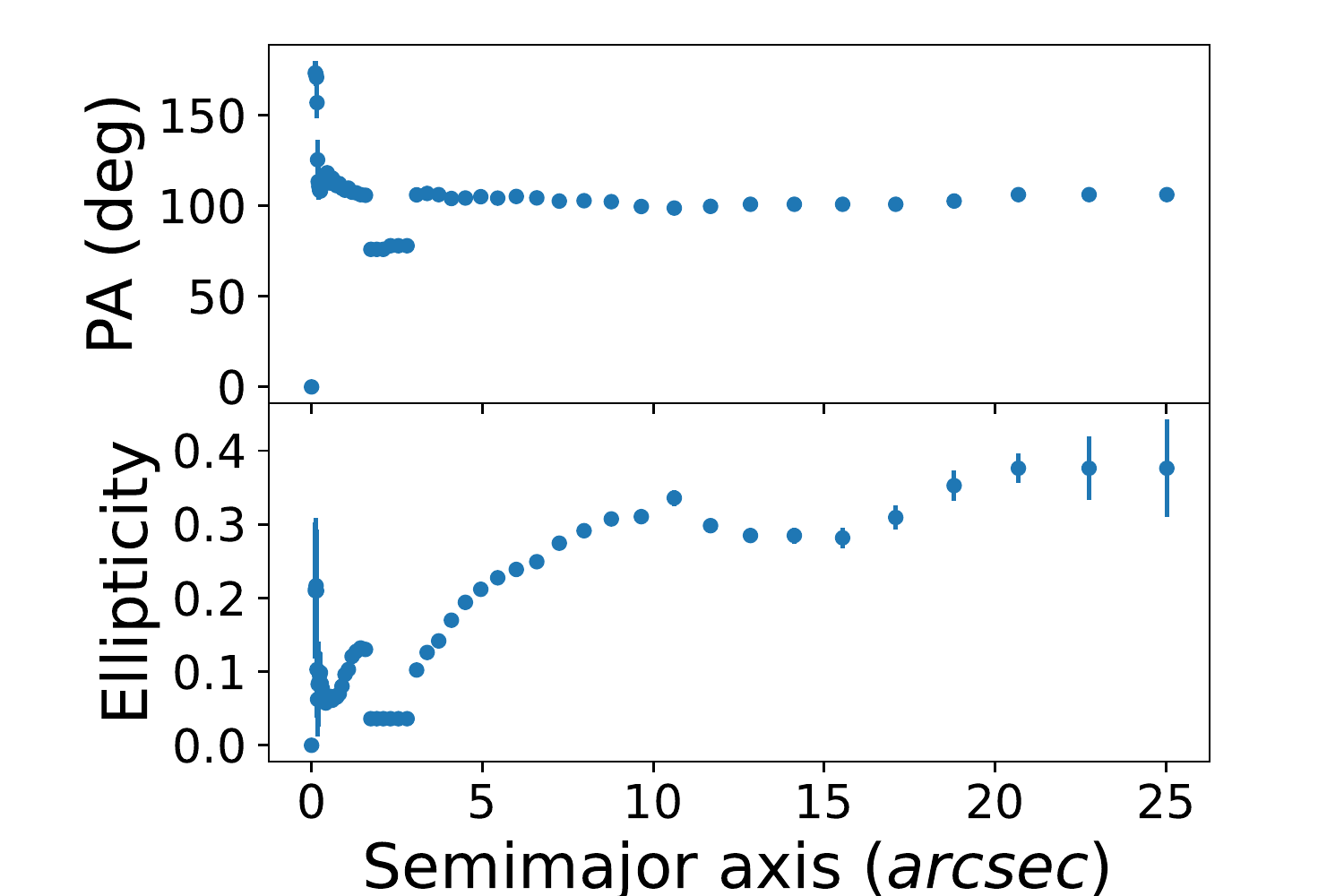} \\

\end{tabular}
\caption {An example of three barred galaxies (G30, G64, G74) and three unbarred galaxies (G78, G139, G364), from our sample showing their contours maps (yellow) and ellipse fitting (red). For each object, the ellipticity and PA of the fitted ellipses as a function of the semi-major axis, are plotted.}
\label{fig:eps_pa}
\end{figure*}

In this work, we have developed an automatic algorithm for classifying barred and unbarred galaxies using ellipse fitting via \textit{photutils.isophote} package in Python, which fits elliptical isophotes to a galaxy image using an iterative method described by \citet{Jedrzejewski1987}. This methodology is advantageous since it can be inplemented and applied to large data sets. We first tested it on a previously defined sample of barred and unbarred galaxies extracted from SDSS-DR7 where an error/misclassification of less than 10$\%$ were found. Accordingly, we have carefully validated our procedure by visual inspection of the V-band images and the explored ellipticity and PA profiles, something that allowed us to assess the reliability of the detection method and to check for false and/or missed identifications.

In our algorithm, we searched for possible signature of a bar feature in the inner two thirds of the semi-major axis (SMA) length, of the ellipticity profile,  after discarding the central three pixels (starting from the center of the galaxy), which is equivalent to $\sim$ 0.6 $\arcsec$ i.e., we excluded the inner 1 to 1.2 kpc. We excluded the central three pixels as errors are significant due to the overlap of elliptical isophotes in this region. With this strategy we will not be able to identify inner or extremely weak short bars, focusing our analysis in strong long bars. Finally, we discarded the analysis on the third part of the SMA as it is unlikely to find a bar in this region where the fluctuations arise probably due to the presence of the spiral arms structure in case of late-type galaxies, or even from perturbations with close neighbors. 

Following this strategy, a galaxy will be defined as barred if the difference between the maximum and minimum values of the ellipticity ($\epsilon_{max}$ and $\epsilon_{min}$, respectively) i.e., $\Delta\epsilon$ $\geq$ 0.1 is achieved with a corresponding variation in the PA that does not exceed 20$^{\circ}$, as customary in similar works \citep{Jogee2004,Aguerri2009a, Barazza2009, Marinova2012, Lansbury2014, Lee2019, Yoon2019}.  Considering that, $\epsilon_{min}$ is defined as the minimum ellipticity after which $\epsilon_{max}$ has been identified. 

In \autoref{fig:eps_pa} we show  six examples of galaxies from our sample (three  barred and three unbarred galaxy) together with their corresponding ellipticity and PA profiles.

Throughout the paper, all the fractions are computed with the respect to the total number of galaxies in our final sample after taking into account the correction of spectroscopic incompleteness.

\section{Results}
\label{sec:results}

We applied both the previously described automated method and a visual inspection on the 3,456 galaxies, of which 906 were identified as barred, corresponding to $\sim26\%$. The largest occurrence was found in the A3880 cluster ($\sim 46.5\%$) and the minimum in A3367 ($\sim14\%$). Our average fraction of barred galaxies is in good agreement with \citet{Barazza2009} who, using a similar bar detection methodology as the one used for this work, studied a sample of field and cluster galaxies with redshift up to 0.8, and found a bar fraction of 24$\%$ and 29$\%$ in the cluster and field galaxies, respectively.

A similar analysis was performed, recently, by \citet{Yoon2019}, exploiting the same bar detection technique, on 105 galaxy clusters at redshift 0.015 $<$ z $<$ 0.060, 16 of which are interacting clusters. They found that the bar fraction in non-interacting clusters is 27$\%$, while the interacting clusters reveal a 1.5 times higher bar fraction (42$\%$). 

Table~\ref{tbl:sample basics} summarizes the basic relevant information for the 32 selected OmegaWINGS clusters including their coordinates, the initial and the selected number of members in each cluster, the fraction of barred galaxies and the cluster X-ray luminosity. 

\begin{table*}
  \centering
   \tabcolsep 5.8pt
   \scriptsize
    \caption{The basic parameters of the 32 OmegaWINGS clusters taken from the OmegaWINGS photometric catalog \citep{Gullieuszik2015}. Column (1) cluster counter (2) indicates the ID of each OmegaWINGS cluster. The third (3) and the fourth (4) columns are the right ascension and the declination sky coordinates in degrees for the cluster centers that coincide with the maximum intensity of the X-ray emission. The fifth (5) column is the initial number of members of galaxies before applying our restrictions. Column (6) is the final number of disk galaxies  after discarding ellipticals and applying our restrictions on surface brightness (SB $<$ 21.5) and ellipticity ($\epsilon$ $<$ 0.5), column (7) represents the bar fraction ($f_{bar}$) in each cluster, column (8) is the X-ray luminosity (in erg s$^{-1}$) of each cluster.}
    \label{tbl:sample basics}
    \begin{tabular}{|r|l|l|l|l|l|l|l|}
\hline
  \multicolumn{1}{|c|}{INDEX} &
  \multicolumn{1}{c|}{ID} &
  \multicolumn{1}{c|}{RA(J2000)} &
  \multicolumn{1}{c|}{Dec(J2000)} &
  \multicolumn{1}{c|}{initial} &
  \multicolumn{1}{c|}{selected} &
    \multicolumn{1}{c|}{$f_{bar}$ ($\%$)} &
    \multicolumn{1}{c|}{$log L_X$ }\\
  \multicolumn{1}{|c|}{(1)} &
  \multicolumn{1}{c|}{(2)} &
  \multicolumn{1}{c|}{(3)} &
  \multicolumn{1}{c|}{(4)} &
  \multicolumn{1}{c|}{(5)} &
  \multicolumn{1}{c|}{(6)} &
    \multicolumn{1}{c|}{(7)} &
    \multicolumn{1}{c|}{(8)}\\
 
\hline
  1 & A85 & 10.4604 & -9.3032 & 172 & 117 & 15 & 44.92 \\
  2 & A151 & 17.2129 & -15.4064 & 165 & 80 & 29 & 44.0 \\
  3 & A168 & 18.7400 & .4309 & 141 & 83 & 33 & 44.04\\
  4 & A193 & 21.2817 & 8.6992 & 72 & 40& 20 & 44.19\\
  5 & A500 & 69.7187 & -22.1108 & 187 & 85 & 23.5 & 44.15\\
  6 & A754 & 137.1350 & -9.6298 & 250 & 138 & 25 & 44.9\\
  7 & A957x & 153.4096 & -.9254 & 48 & 24 & 21 & 43.89 \\
  8 & A970 & 154.3571 & -10.6889 & 136 & 64 & 27& 44.18  \\
  9 & A1631a & 193.2192 & -15.4133 & 288 & 152 & 27& 43.86 \\
  10 & A2382 & 327.9817 & -15.7059 & 271 & 132 & 26 & 43.96\\
  11 & A2399 & 329.2571 & -7.8394 & 234 & 114 & 26 & 44.0 \\
  12 & A2415 & 331.4108 & -5.5922 & 144 & 70 & 24 & 44.23 \\
  13 & A2457 & 338.9200 & 1.4850 & 232 & 114 & 21 & 44.16 \\
  14 & A2717 & 0.8042 & -35.9370 & 135 & 60 & 32& 44.0\\
  15 & A2734 & 2.8400 & -28.8543 & 220 & 107 & 21& 44.41\\
  16 & A3128 & 52.4608 & -52.5797 & 333 & 176& 28.5& 44.33\\
  17 & A3158 & 55.7208 & -53.6313 & 243 & 118 & 26 & 44.73\\
  18 & A3266 & 67.8054 & -61.4533 & 479 & 195 & 22.5 & 44.79\\
  19 & A3376 & 90.1712 & -40.0446 & 229 & 107  & 23 & 44.39\\
  20 & A3395 & 96.9012 & -54.4494 & 244 & 122 & 21& 44.45\\
  21 & A3528 & 193.8381 & -28.9505 & 262 & 120 & 31 & 44.12\\
  22 & A3530 & 193.9000 & -30.3476 & 275 & 126 & 35 & 43.94\\
  23 & A3532 & 194.3417 & -30.3636 & 107 & 34 & 26.5 & 44.45\\
  24 & A3556 & 201.0279 & -31.6699 & 328 & 134 & 28 & 43.97\\
  25 & A3558 & 201.9867 & -31.4956 & 442 & 155& 33.5 & 44.8\\
  26 & A3560 & 202.9729 & -33.2342 & 244 & 123 & 32 & 44.12\\
  27 & A3667 & 303.1137 & -56.8268 & 386 & 169 & 14 & 44.94\\
  28 & A3716 & 312.8329 & -52.6362 & 327 & 132& 26 & 44.0\\
  29 & A3809 & 326.7462 & -43.8989 & 189 & 91& 20 & 44.35\\
  30 & A3880 & 336.9767 & -30.5755 & 216 & 99& 46.5 & 44.27\\
  31 & A4059 & 359.2529 & -34.7591 & 229 & 108& 34 & 44.49\\
  32 & IIZW108 & 318.4829 & 2.5654 & 162 & 67 & 16 & 44.34\\
\hline\end{tabular}
\end{table*}

\begin{figure*}
\centering
\begin{tabular}{lll}
\includegraphics[width=0.3\textwidth]{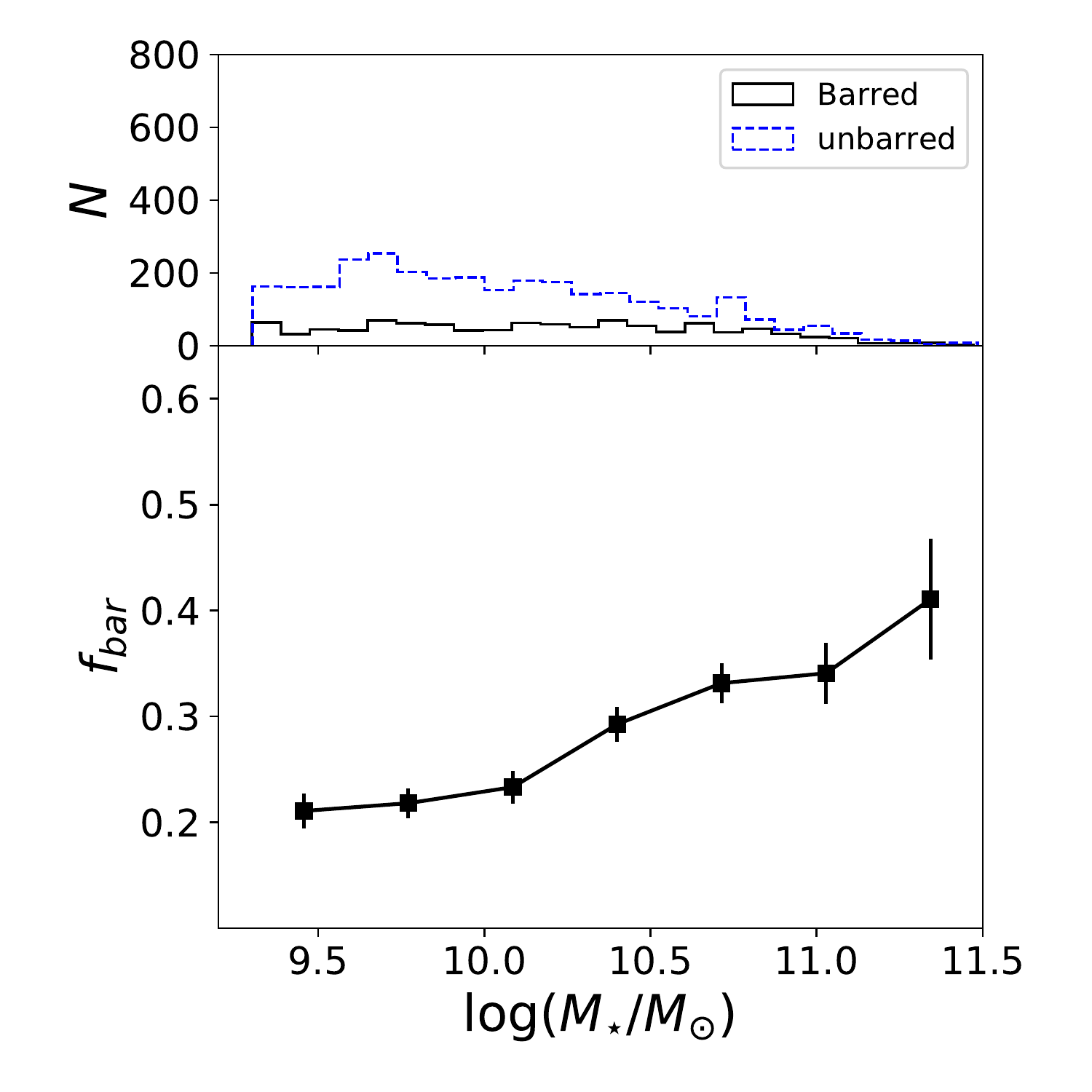} &
\includegraphics[width=0.3\textwidth]{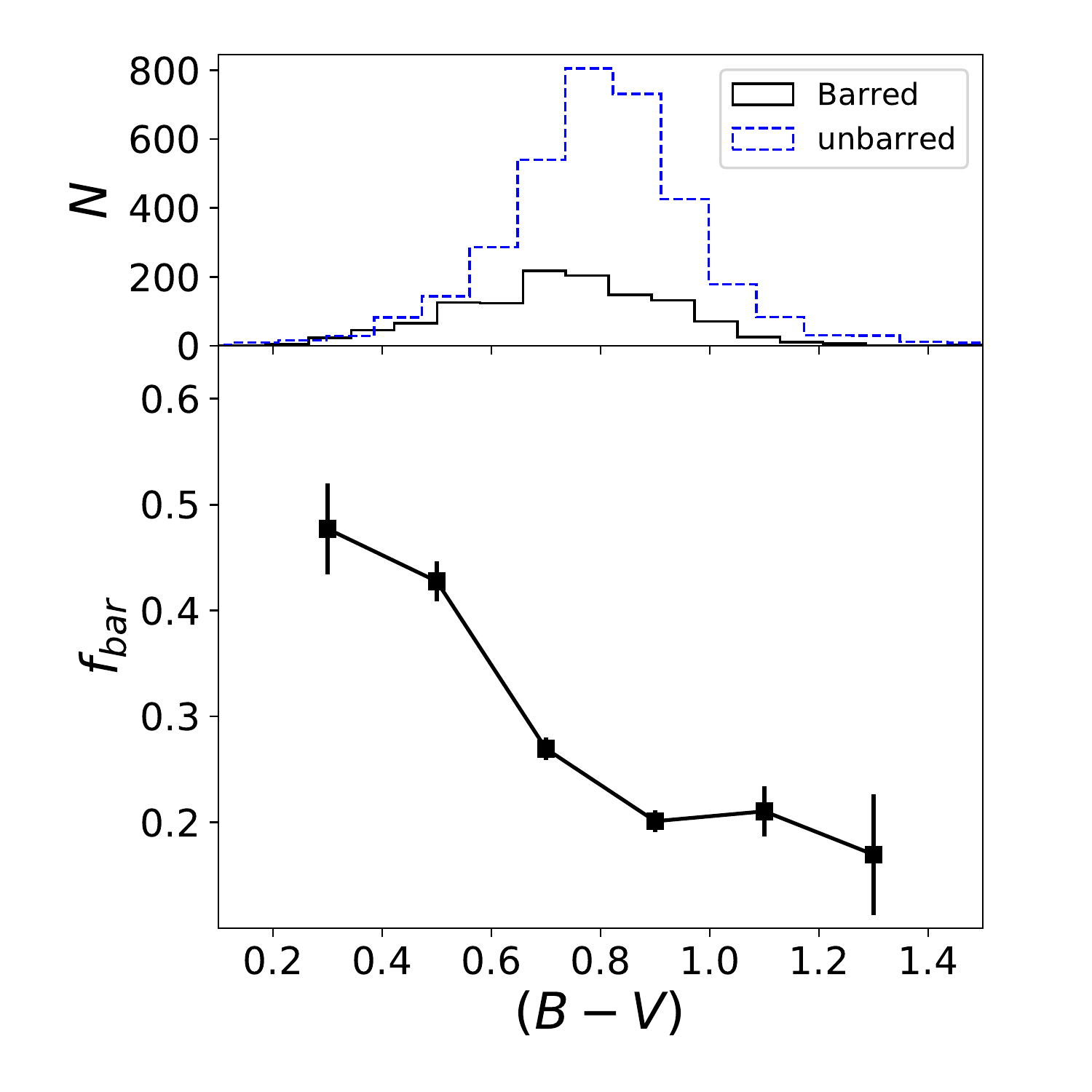} &
\includegraphics[width=0.3\textwidth]{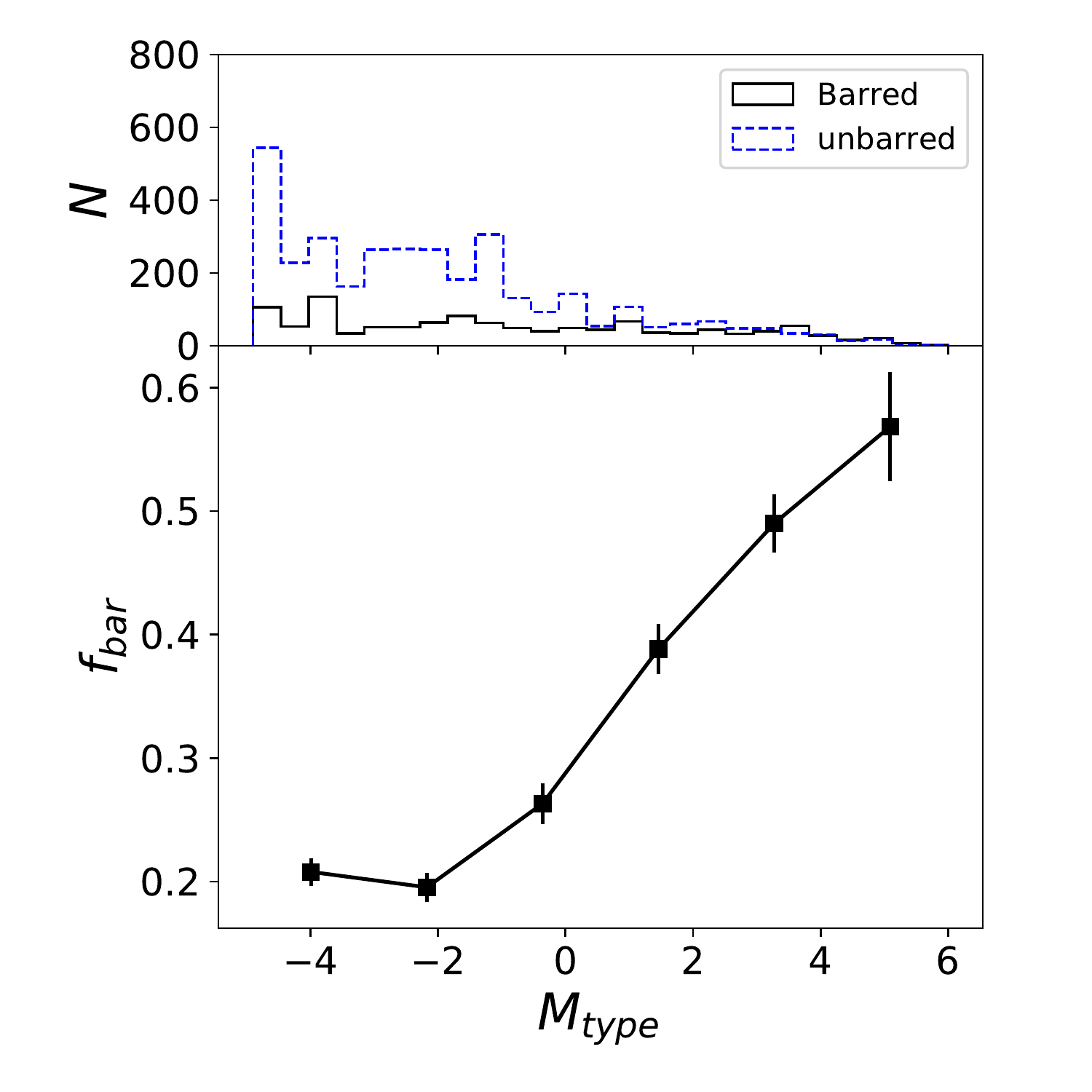} \\
\includegraphics[scale=0.9, bb=10 10 50 170]{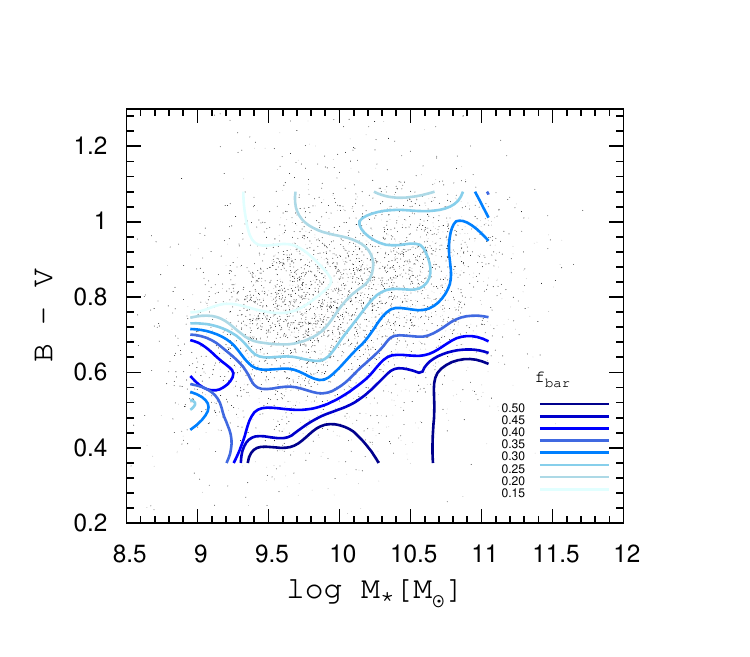} &
\includegraphics[scale=0.9, bb=10 10 50 170]{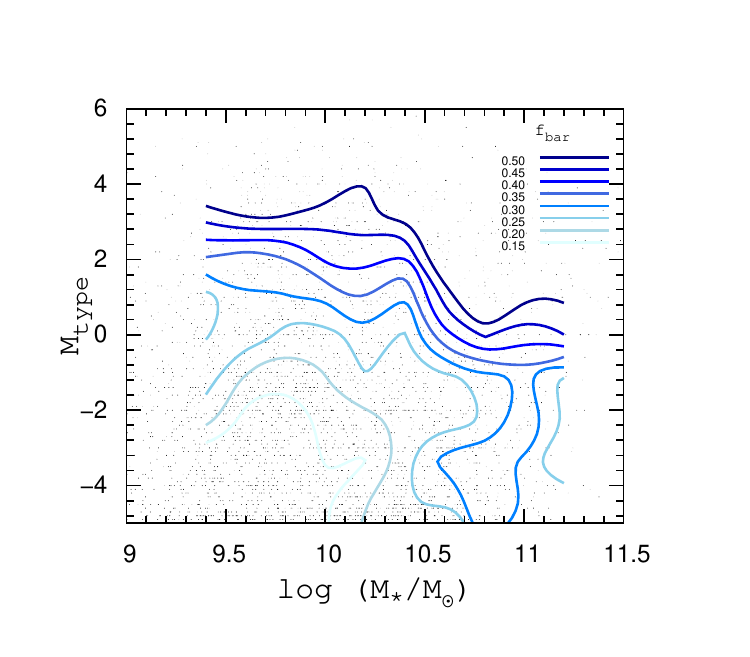} &
\includegraphics[scale=0.9, bb=10 10 50 170]{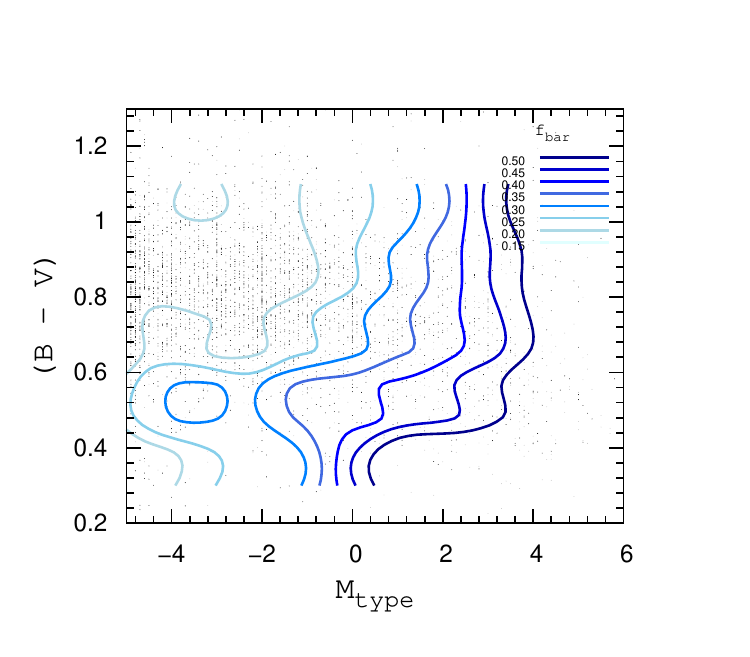} \\
\end{tabular}
\caption{\textit{Top panels:} Fraction of barred galaxies  $f_{bar}$ as a function of stellar mass (log($M_{\star}/M_{\odot}$), left panel), color ((B-V), middle panel) and morphological type (M$_{type}$, right panel). Error bars indicate the estimated 1$\sigma$ confidence intervals based on the bootstrapping resampling method. Also included are the distributions of each property for the case of barred (solid black line) and unbarred (segmented blue line) galaxies. \textit{Bottom panels:} Bar fraction $f_{bar}$ isocontours in the $(B-V)$ vs (log($M_{\star}/M_{\odot}$) (left panel), M$_{type}$ vs (log($M_{\star}/M_{\odot}$) (middle panel) and $(B-V)$ vs M$_{type}$ (right panel) spaces. Contours denote regions of constant $f_{bar}$ in the range 0.15 $\leq f_{bar} \leq$ 0.50.}
\label{fig:fbar_galpar}
\end{figure*}

\subsection{Dependence of the fraction of barred galaxies on intrinsic galaxy properties}
We start our analysis by looking at the dependence of the bar fraction $f_{\rm bar}$ as a function of three relevant  parameters; the total stellar mass (M$_{*}$), the $(B-V)$ color, and the morphological type (M$_{\rm type}$). Error bars in all figures indicate the estimated 1$\sigma$ confidence intervals based on the bootstrapping resampling method over 100 realizations. 

In Figure~\ref{fig:fbar_galpar} top left panel, we show the well known trend of increasing bar fraction with increasing total stellar mass, a trend usually found when studying strong bars \citep{Masters2012, Skibba2012,Cervantes2013, Gavazzi2015}, as is the case of our sample. This trend is expected if we consider that bars form earlier in massive galaxies, once they build a dynamically cold disk, prone to experience gravitational instabilities that trigger the formation of the bar \citep{Sheth2012, Saha2018}.

Recently, \cite{Yoon2019} performed a study on 105 galaxy clusters, selected from the Sloan Digital Sky Survey, and found  the bar fraction is higher towards massive galaxies, a pertinent comparison with our result, given that both samples encompass cluster galaxies and the bar detection is performed in a similar way.

Figure~\ref{fig:fbar_galpar}, top central panel, illustrates $f_{bar}$ as a function of the rest-frame (B-V) color, reveling a strong negative trend, with a significant decrease in the bar occurrence for increasing (B-V) values. Our finding seems to contradict previous studies \citep{Masters2011, Lee2012} that report a higher likelihood of finding bars in red galaxies. One possible explanation of this difference, is that our sample consists of cluster galaxies, and this environment is known to produce strong transformations both in stellar populations and in morphology. To explore this further,  Figure~\ref{fig:fbar_galpar} top right panel presents the fraction of barred galaxies as a function of morphological type with a clear increase of $f_{\rm bar}$ as we go from early- to late-type galaxies. The dependence of $f_{\rm bar}$ on morphology is one that is expected, as bars require a dynamically cold disk to form and grow \citep{Sheth2012, Saha2018}. 

Finally, regarding the trend of the bar fraction with morphology in cluster galaxies, \citet{Yoon2019} also found that the fraction of barred galaxies decreases with increasing bulge-to-total light ratio, in line with our result, taking into account that this ratio decreases with increasing M$_{type}$.

Having identified the strong dependence of the bar fraction on these three parameters,  we also need to take into account that they are not mutually independent.  To explore this aspect, we analyze the co-dependence of the bar fraction on the combination of two of these three parameters by looking at the bar fraction in two dimensional parameter spaces in the bottom panels of  \autoref{fig:fbar_galpar}, in which the distribution of $f_{bar}$ is smoothed following the same technique used by \citet{Lee2012} and \citet{Cervantes2013} through a cubic B-spline kernel, and calculating the ratio of the weighted number of barred galaxies to the total number of weighted galaxies using a fixed-size smoothing scale. 

We present the bar fraction as a function of stellar mass and color, with a clear increase of the bar fraction for more massive, bluer galaxies. The contours of fixed bar fraction indicate a co-dependence on both parameters, as already evident in the left and central upper panel of the same figure, but exploring the bar fraction in the (B $-$ V) vs. M$_{\rm type}$ space (bottom right panel of  \autoref{fig:fbar_galpar}), we find that the dependence on color vanishes once we fix the morphological type. This indicates that the dependence of $f_{\rm bar}$ on (B$-$V) is mostly driven by morphology, except for the bluest galaxies, and explains why we find a higher bar fraction in bluer galaxies, as they are preferentially late-types.

For completeness, we also explore the bar fraction in the M$_{\rm type}$ vs. $M_{\star}$ space, finding a constant bar fraction at fixed M$_{\rm type}$ for late-type galaxies with $M_{\star} < 10^{10.5}M_{\odot}$, and an increase of $f_{\rm bar}$ for more massive galaxies.

Once identified the total stellar mass and the morphological type as two determinant parameters in determining the likelihood of finding barred galaxies in our sample, we turn to study the influence of the environment in the following sections.

\subsection{Bar fraction and the cluster environment}

\begin{figure}
\centering
\includegraphics[width=0.4\textwidth]{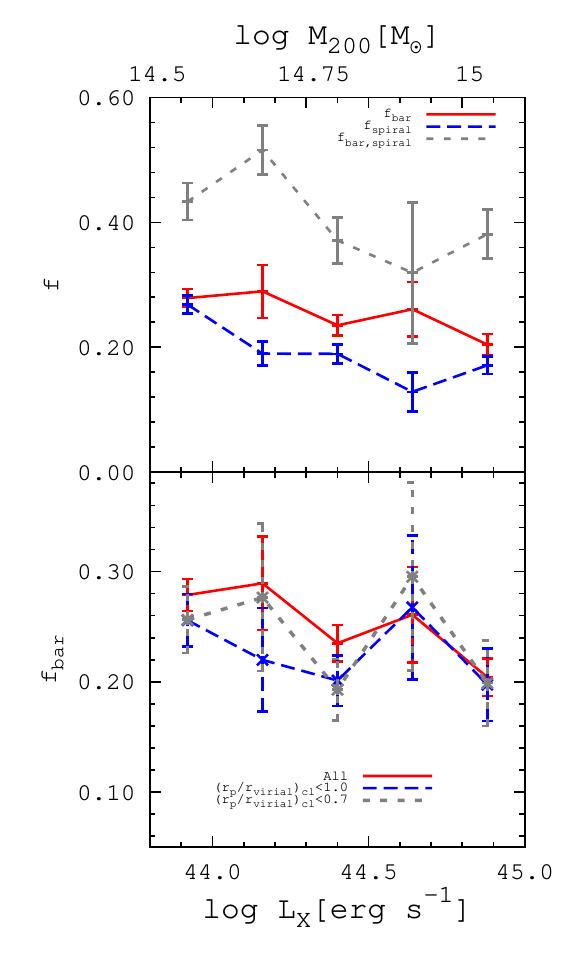}
\caption{ \textit{Top panel:} Fraction of barred ($f_{bar}$, red solid line), spiral ($f_{spiral}$, blue segmented line) and barred spiral ($f_{bar,spiral}$, gray dashed line) galaxies as a function of X-ray luminosity and total dynamical mass. \textit{Bottom panel:} Fraction of barred galaxies as a function of X-ray luminosity and total dynamical mass, for the full sample (red solid line), and for galaxies within 1.0 (blue segmented line) and 0.7 (gray dashed line) virial radii.}
\label{fig:LX_fbar}
\end{figure}

We start by exploring the dependence of the fraction of barred galaxies on the global properties of the clusters in the sample. Figure~\ref{fig:LX_fbar} shows a decrease of the fraction of barred galaxies with increasing X-ray luminosity, here used as a proxy for the cluster mass \citep{Reiprich02}, indicating that barred galaxies are less abundant in massive galaxy clusters. Having identified the morphological type as a key feature to establish the likelihood of a galaxy hosting a bar, and given that the fraction of early-type galaxies is higher in more massive clusters \citep{poggianti09} and confirmed by P\'erez-Mill\'an et al., submitted, we also explore the fraction of spiral galaxies ($M_{type} >1$) as a function of X-ray luminosity, in order to determine if the decline of barred galaxies in massive clusters could be attributed to a decline of the late-type population. The fraction of spiral galaxies and the bar fraction among the spiral sample as a function of X-ray luminosity are also shown in Figure~\ref{fig:LX_fbar}, presenting the expected decline when moving from low to high X-ray luminosity, supporting our scenario in which the presence of a bar is mostly determined by morphology.

It is important to point out that this result relies on the analysis of the full sample as described in Section 3, with an underlying bias due to the FOV of the survey that systematically covers areas farther beyond the virial radius for clusters with smaller projected sizes. If we instead control our analysis by imposing a limiting virial radius for the estimation of $f_{bar}$, the trend with the X-ray luminosity weakens as is the case for the analysis restricted for galaxies within 1.0 and 0.7 virial radii (bottom panel of Figure~\ref{fig:LX_fbar}). 

\begin{figure*}
\centering
\subfloat{\includegraphics[width=0.4\textwidth]{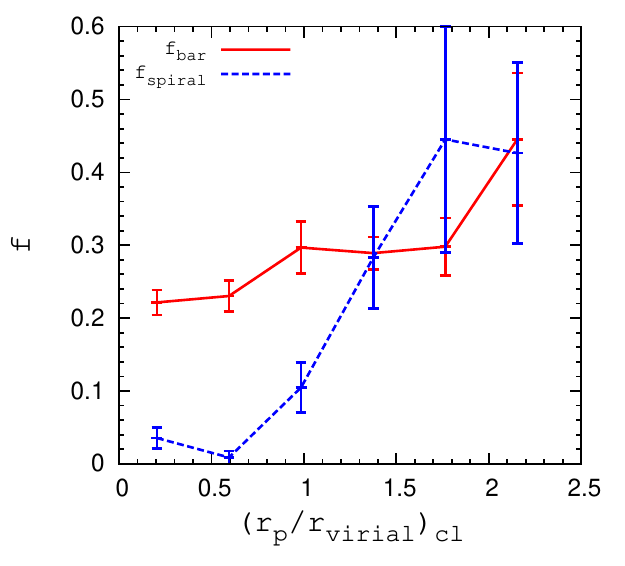}}
\subfloat{\includegraphics[width=0.45\textwidth]{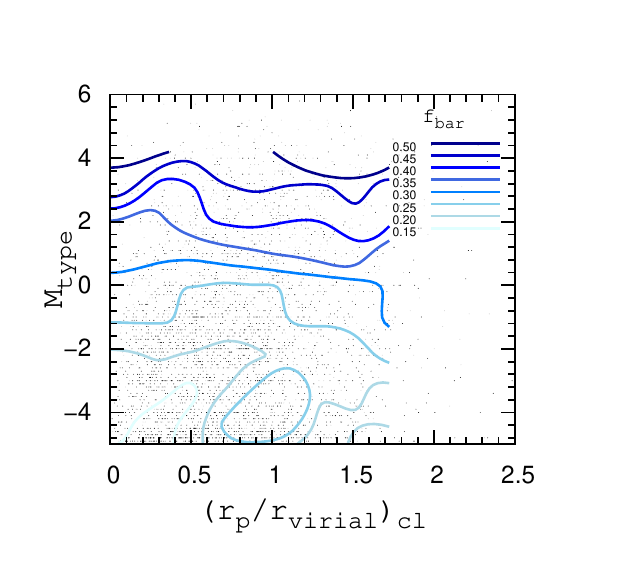}}\\
\caption{\textit{Left panel:} The fraction of barred ($f_{bar}$) and spiral ($f_{spiral}$) galaxies as a function of the projected clustercentric distance, normalized to the virial radius of the cluster ($(r_{p}/r_{virial})_{cl}$). \textit{Right panel:} Bar fraction isocontours in the M$_{type}$ vs $(r_{p}/r_{virial})_{cl}$ space. Contours denote regions of constant $f_{bar}$ in the range 0.15 $\leq f_{bar} \leq$ 0.50.}
\label{fig:fbar_rp}
\end{figure*}

Turning to the distribution of barred galaxies within clusters, left panel of \autoref{fig:fbar_rp}, presents the bar fraction as a function of the normalized projected clustercentric distance ($(r_p/r_{virial})_{cl}$), showing a decrease of the bar fraction for decreasing  distances, with a drastic drop once we cross the virial radius of the clusters, where the bar fraction is reduced to half of the value found among galaxies in the outskirts. A possible explanation, at least partial, to this trend, can be given by considering the so called morphology-density relation \citep[MDR,][]{dressler80} combined to  the occurrence of bars as a function of the host galaxy morphology. For our restricted sample of face-on, S0 to late-type galaxies, the corresponding morphology-density relation is shown in the left panel of \autoref{fig:fbar_rp} (blue dashed line), recovering the expected decline of $f_{spiral}$ to the center of the clusters.

To further explore whether the dependence of the bar fraction with the projected clustercentric distance is merely a reflection of the morphology-density relation, in the right panel of \autoref{fig:fbar_rp} we analyze the behavior of $f_{bar}$ in the 2D space of morphological type and projected clustercentric distance. For late-type galaxies ($M_{\rm type} >$ 1), at fixed morphological type, the bar fraction remains constant as a function of clustercentric distance. It is only for S0 galaxies ($M_{\rm type} < $ 0), that even at fixed morphological type we can still detect a dependence on the projected clustercentric distance, with $f_{bar}$ decreasing once we cross the virial radius of the cluster. The decrease of $f_{\rm bar}$ among S0 galaxies for decreasing clustercentric distance could be a reflection of the infall time experienced by the different galaxies, with the ones still retaining their bars in the outer parts of the clusters with the most recent infall times and still experiencing a morphological transformation. 

\begin{figure}
\centering
\includegraphics[scale=1.3, bb=10 10 200 180]{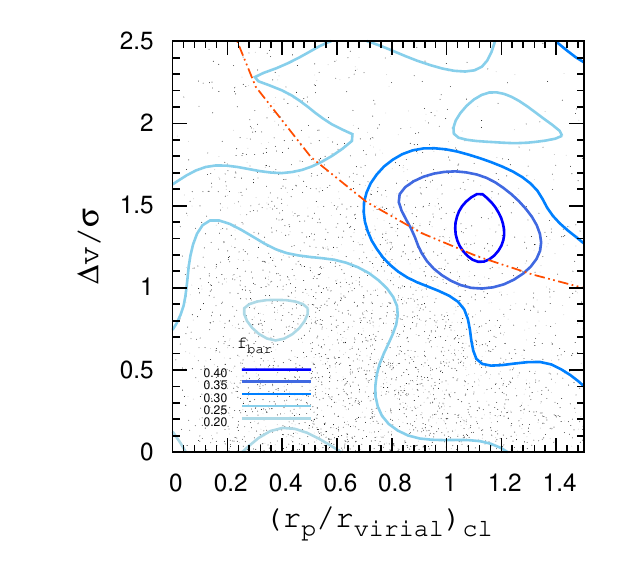}
\caption{Bar fraction isocontours in the $\Delta v/\sigma$ vs $(r_{p}/r_{virial})_{cl}$ space. Contours denote regions of constant $f_{bar}$ in the range 0.20 $\leq f_{bar} \leq$ 0.40. The red dash-dotted line corresponds to the escape velocity in a NFW halo.}
\label{fig:fbar_rp_v}
\end{figure}

A more complete picture of a possible environmental cluster effect can be gained by analyzing the occurrence of bars in a phase-space diagram, relating the normalized projected clustercentric distance and the line-of-sight velocity normalized by the cluster velocity dispersion ($\Delta(v)/\sigma$), where galaxies trace a typical path to settle into the cluster potential. The corresponding phase-space diagram for our sample is shown in \autoref{fig:fbar_rp_v}, where the contours denote the bar fraction and the red dash-dotted line corresponds to the escape velocity in a Navarro, Frenk and White \citep[NFW,][]{Navarro1996} halo with a concentration $c=6$ for reference. At infall, galaxies that approach the center of the cluster are gaining velocity by the effect of the cluster potential. As the galaxies experience dynamical friction, repeated fast interactions with members of the cluster, and violent relaxation, they loose velocity concentrating in the lower left region of the diagram. As can be noted by the contours, the highest occurrence of bars is at large clustercentric distances and high velocities, with a peak of the bar fraction just above the escape velocity demarcation line, an indication that barred galaxies are recent infallers. Furthermore, the contour denoting $f_{bar}=0.30$ presents an elongated shape with increasing phase-space velocity with decreasing clustercentric distance, as would be expected for recent infalling galaxies \citep{Rhee2017, Jaffe2019}. 

\subsection{The influence of the nearest neighbor galaxy}
Galaxies in clusters experience a variety of environmental processes that in principle could promote, suppress, or even destroy bars. In the previous section, we explored the dependence of the bar fraction as a function of total cluster mass and as a function of clustercentric distance, but within clusters galaxies also experience galaxy-galaxy interactions. In order to investigate if this kind of interaction influences the occurrence of bars in the cluster environment, we made a similar analysis as followed by \citet{park09}, measuring the projected distance to the other galaxies, and determining the closest one. This distance is normalized by the virial radius of the neighbor $(r_{\rm virial})_{\rm neigh}$, calculated as the radius of a sphere whose density of baryonic matter (approximated by only taking into account the stellar mass obtained through {\sc sinopsis}) is equal to 200 times the critical density of the universe at that redshift:
\begin{equation}
    R_{\textrm{vir}}=\left(\frac{3}{4\pi}\cdot \frac{M_\star}{200 \rho_{\rm c}}\right)^{1/3}.
\end{equation}

\begin{figure}
\centering
\begin{tabular}{l}
\includegraphics[width=0.43\textwidth]{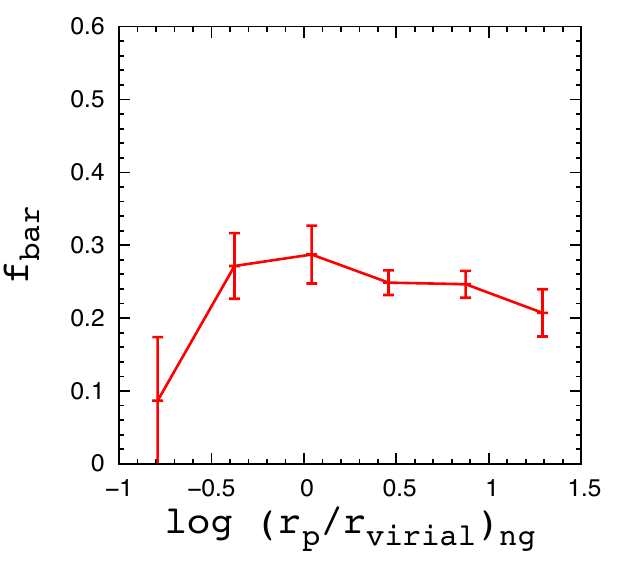}\\
\includegraphics[width=0.43\textwidth]{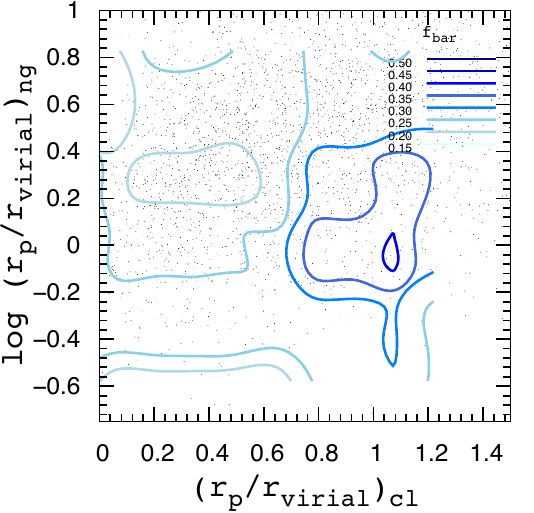}
\end{tabular}
\caption{\textit{Top panel:} The fraction of barred galaxies as a function of the distance to the nearest neighbor, normalized to the virial radius of the neighbor. \textit{Bottom panel:} Bar fraction isocontours in the log ($r_p/r_{virial}$)$_{ng}$ vs $(r_{p}/r_{virial})_{cl}$ space. Contours denote regions of constant $f_{bar}$ in the range 0.15 $\leq f_{bar} \leq$ 0.50.}
\label{fig:Np}
\end{figure}

This strategy was applied only on galaxies spectroscopically confirmed as cluster members (P\'erez-Mill\'an et al., submitted). With these results, we compute the bar fraction as a function of the distance to the nearest neighbor galaxy, as shown in the top panel of Figure~\ref{fig:Np}, where we note a fairly constant $f_{bar}$ in galaxies with a separation larger than one virial radius. But when we look at close pairs, log ($r_p/r_{virial}$)$_{ng}$ $<$ -0.4, the fraction of barred galaxies presents a dramatic decline. This result confirms previous findings indicating that close gravitational encounters between galaxies are able to perturb galaxies to a point that bars are destroyed. Such results were for example reported by \citet{Lee2012} and \citet{Lin2014} for the general galaxy field population in the Local Universe.

Figure~\ref{fig:Np} bottom panel presents the co-dependence of the bar fraction on the clustercentric distance and the distance to the nearest neighbor. It is clear that the maximum $f_{bar}$ value are found in galaxies located beyond the virial radius of the clusters, i.e not yet experiencing its potential well, with increasing bar fraction for increasing clustercentric and neighbor distances, reaching a maximum at the virial radii in both axis. The maximum of $f_{bar}$ at this precise location might indicate a ``sweet spot'' where the gravitational potential of the cluster and/or the nearest neighbor triggers the formation of the bar \citep{Romano2008,Lokas2016,Martinez2016, Lokas2020} while not being strong enough to dissolve preexisting ones \citep{Lee2012, Lin2014, Zana2018}.

\subsection{The relation between the presence of the bar and the central star formation activity}

Using the results from {\sc sinopsis}, we study how the presence of a bar might affect the properties of the stellar populations, possibly influencing the ability of a galaxy to boost/enhance or quench the star formation in the central parts.

One way to study this aspect, is to analyze possible dependencies of the specific star formation rate (sSFR), i.e. the star formation rate per unit stellar mass, on the presence/absence of bars. We do this by considering the SFR averaged on a $5.6\times 10^8$ year age bin (this corresponds to the average of SFR$_1$ and SFR$_2$, see \autoref{sec:properties}). Using such a large bin has a twofold advantage with respect to considering the more commonly used $\sim 10^7$ years bin, that is the one which is characterized by the presence of emission lines in the spectra. Firstly, it allows us to use a much larger sample, as the cluster galaxies population is rich in quenched ones (hence, displaying no emission lines). Secondly, and maybe most importantly, sampling a longer age allows to average effects that are local in time, as for example star formation quenching or triggering by external, rapid mechanisms such as ram pressure \citep{vulcani18} or neighbors encounter \citep{park09}, which are both expected to act on relatively short timescales. A stellar bar, on the other side, is a morphological feature that is much more long-lived and whose possible effects are hence better studied on longer time scales.

We try to isolate the effects of the bar on the sSFR, by contrasting them against the stellar mass, the morphological type and the projected clustercentric distance normalized to the cluster virial radius. 

\begin{figure*}[!t]
\centering
\begin{tabular}{ccc}
\includegraphics[scale=0.9, bb=20 10 200 170]{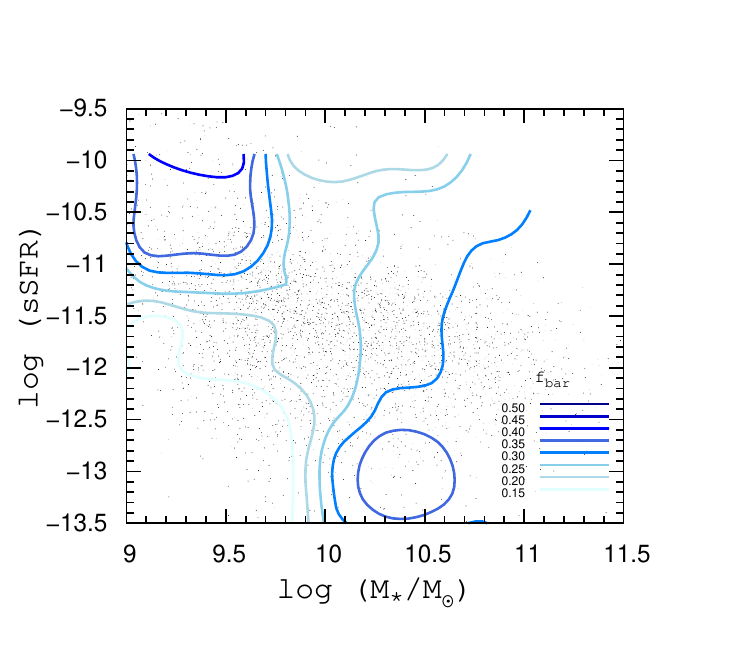} & 
\includegraphics[scale=0.9, bb=20 10 200 170]{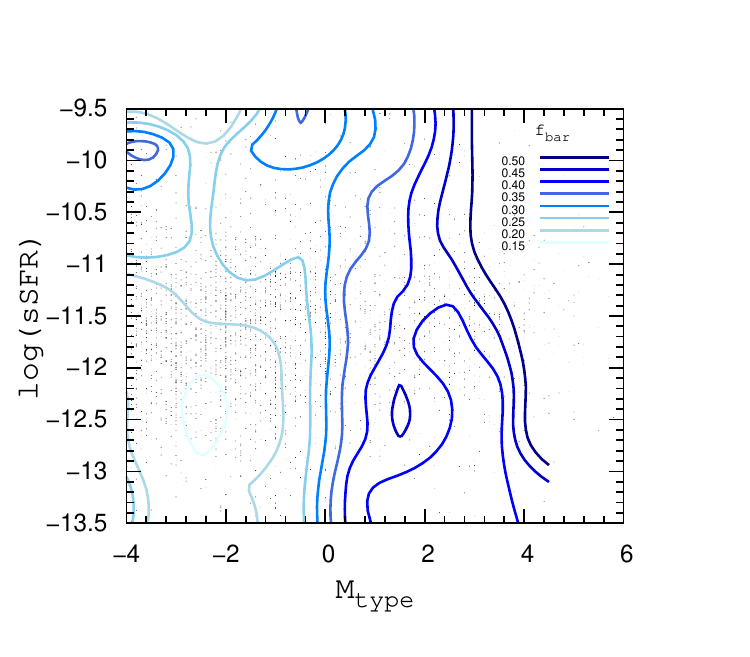} & 
\includegraphics[scale=0.9, bb=20 10 200 170]{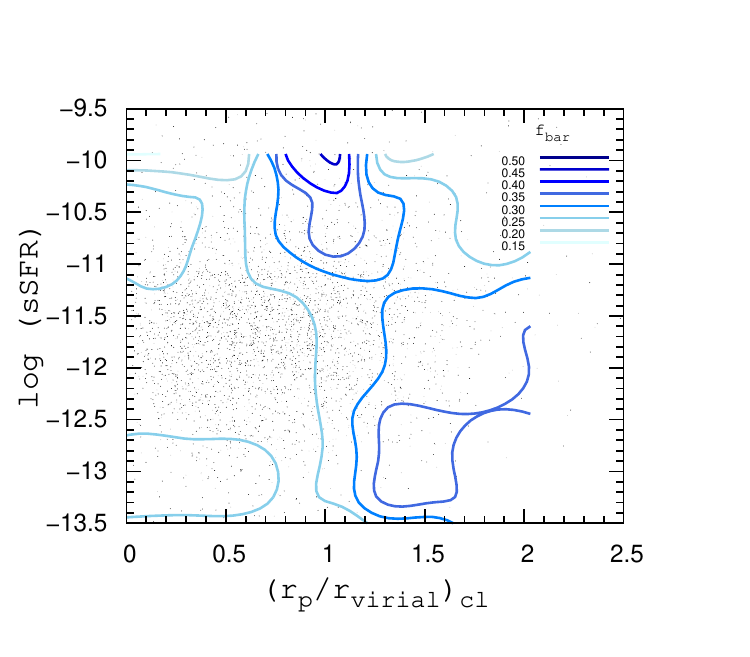} \\ 
\end{tabular}
\caption{Bar fraction isocontours in the log (sSFR) vs log ($M_{\ast}/M_{\odot}$) (left panel), M$_{type}$ (middle panel) and $(r_{p}/r_{virial})_{cl}$ (right panel) spaces. Contours denote regions of constant $f_{bar}$ in the range 0.15 $\leq f_{bar} \leq$ 0.50.}
\label{fig:SFR_Mass}
\end{figure*}

In the leftmost panel of \autoref{fig:SFR_Mass}, we analyze the occurrence of stellar bars in the context of the galaxy main sequence, i.e. the relation between stellar mass and sSFR. Two regions clearly stand out in this plot: high sSFR and low masses, and low sSFR and intermediate-high masses, both characterized by the highest bar fraction. Before adventuring in an interpretation of this finding, we need to take into account possible effects coming from the  mass-morphology relation. In the central panel of \autoref{fig:SFR_Mass} the relation between sSFR and morphology is displayed, again with the barred galaxy fractions as a third dimension. Here, once again, the importance of morphology is evident: an almost monotonic trend is observed, at basically all sSFR values, of increasing bar fraction as a function of the morphological type. 

Other two aspects that need to be considered in this context are the morphology-density relation (MDR), and the fact that dense environments tend to quench star formation. Because of the MDR in particular, we might be led to suggest that the cluster is promoting (inhibiting) bar destruction (creation), while evidence we have so far accumulated suggests instead that the bar fraction has a strong dependence with morphology which, in turns, depends on the local environment. We can explore this in the rightmost panel of the same figure, where we look for the highest concentration of barred galaxies in the sSFR vs projected clustercentric distance space. As already observed for the main sequence relation, here as well we find two areas in the parameter space where barred galaxies mostly concentrate: an area of high sSFR, likely corresponding to the low mass end (see the leftmost panel), and an area at the lowest sSFR. Barred galaxies with the highest sSFR values are found among those that are close to the virial radius of the clusters: this is where bars seem to provide the strongest effect in promoting star formation. On the other hand, the fact that the majority of barred galaxies with high sSFR have low stellar mass, might suggest that this is where the gas reservoir is the largest. In fact, if we look at the region of the plot in which the other peak of barred galaxies is located, at low sSFR values (hence, again by comparison of the leftmost panel, the higher mass galaxies), these are most likely galaxies that, being even further away from the cluster center, have not started yet a morphological transformation and are, hence, more likely to host a stellar bar. These are likely to be more massive galaxies that, probably due to pre-processing before in-falling, have consumed most of their gas and are displaying lower levels of star formation. Also, the interaction with the cluster environment is not effective yet, at these distances, and the influence of the bar in promoting the SFR might hence be limited.

\begin{figure}
\centering
\begin{tabular}{c}
\includegraphics[width=0.4\textwidth]{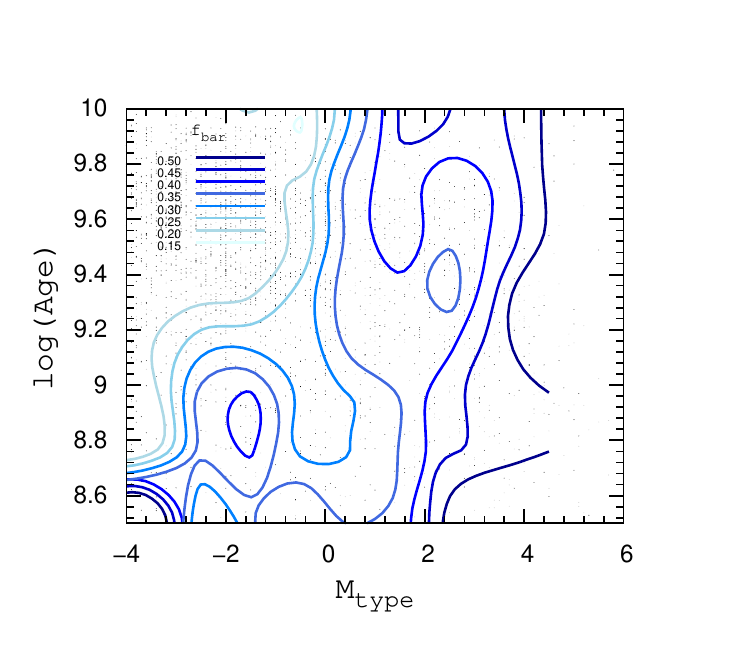} \\  
\includegraphics[width=0.4\textwidth]{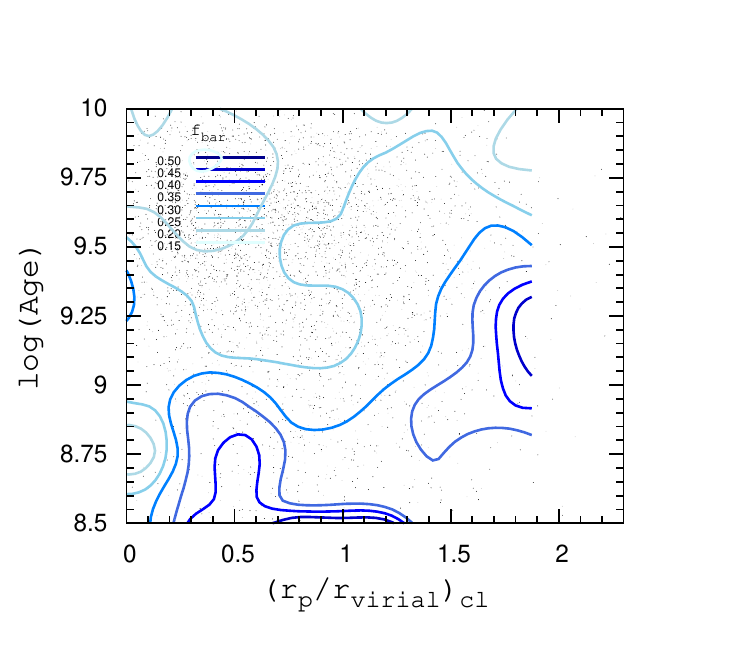} \\  
\end{tabular}
\caption{Bar fraction isocontours in the log (Age) vs  M$_{type}$ (top panel) and $(r_{p}/r_{virial})_{cl}$ (bottom panel) spaces. Contours denote regions of constant $f_{bar}$ in the range 0.15 $\leq f_{bar} \leq$ 0.50.}
\label{fig:SSPage}
\end{figure}

The mix of stellar populations in a galaxy can be globally parametrized by the average galaxy age, and is the result of its star formation history. In the upper panel of \autoref{fig:SSPage}, we analyze the occurrence of barred galaxies in the luminosity-weighted age vs morphological type plane. Again, a strong dependence of the barred fraction on the morphological type is found, with a fairly constant trend as a function of the morphological type. On top of this, we observe a slight tendency in barred early spirals of being younger than their later types counter parts. A secondary peak in the bar fraction is observed at the youngest ages and earliest morphologies, an indication that bars tend to ``centrally rejuvenate'' galaxies in the cluster environment. This result is in line with the reported bar rejuvenation in S0 galaxies by \citet{Barway2020} who found bluer bars in S0 galaxies when compared with spiral galaxies, attributing this difference in color to induced star formation activity in S0 galaxies triggered by mergers or tidal interactions in intermediate-density environments.

In the lower panel of the same figure, we analyze the bar fraction in the plane of stellar age and clustercentric distance, the latter being used as a broad proxy of the intensity of the cluster influence on the galaxies. Again, two peaks in the bar fraction distribution are found: the first is located at larger distances and intermediate ages. The latter are late type galaxies, that are barely in-falling into the cluster, and that are still retaining their morphologies, and in our view they would hence display the highest probability of hosting a stellar bar. The other peak is found at around 0.5-1 R$_{\mbox{vir}}$, and at the youngest ages. These are likely in their first interaction with the cluster, for which the presence of a bar combined to other effects such as ram-pressure, can favor central star formation activity, visible through younger average ages.

Another quite effective and more general mean of quantifying the influence of a stellar bar on the properties of the stellar populations, can be achieved by using ``quenching indexes'' (see Pérez-Millán et al., submitted), that is the ratios between SFR in two contiguous age bins as defined in \autoref{sec:sample}:
\begin{equation}
    S_{i,j}=\frac{SFR_i}{SFR_j},    
\end{equation}
in which the index $i$ refers to an older age bin than $j$. Like this, a quenching index value larger than 1 represents a diminishing SFR, while an enhancement is expected for values lower than 1.

\begin{figure*}[!ht]
\centering
\begin{tabular}{ccc}
\includegraphics[width=0.33\textwidth]{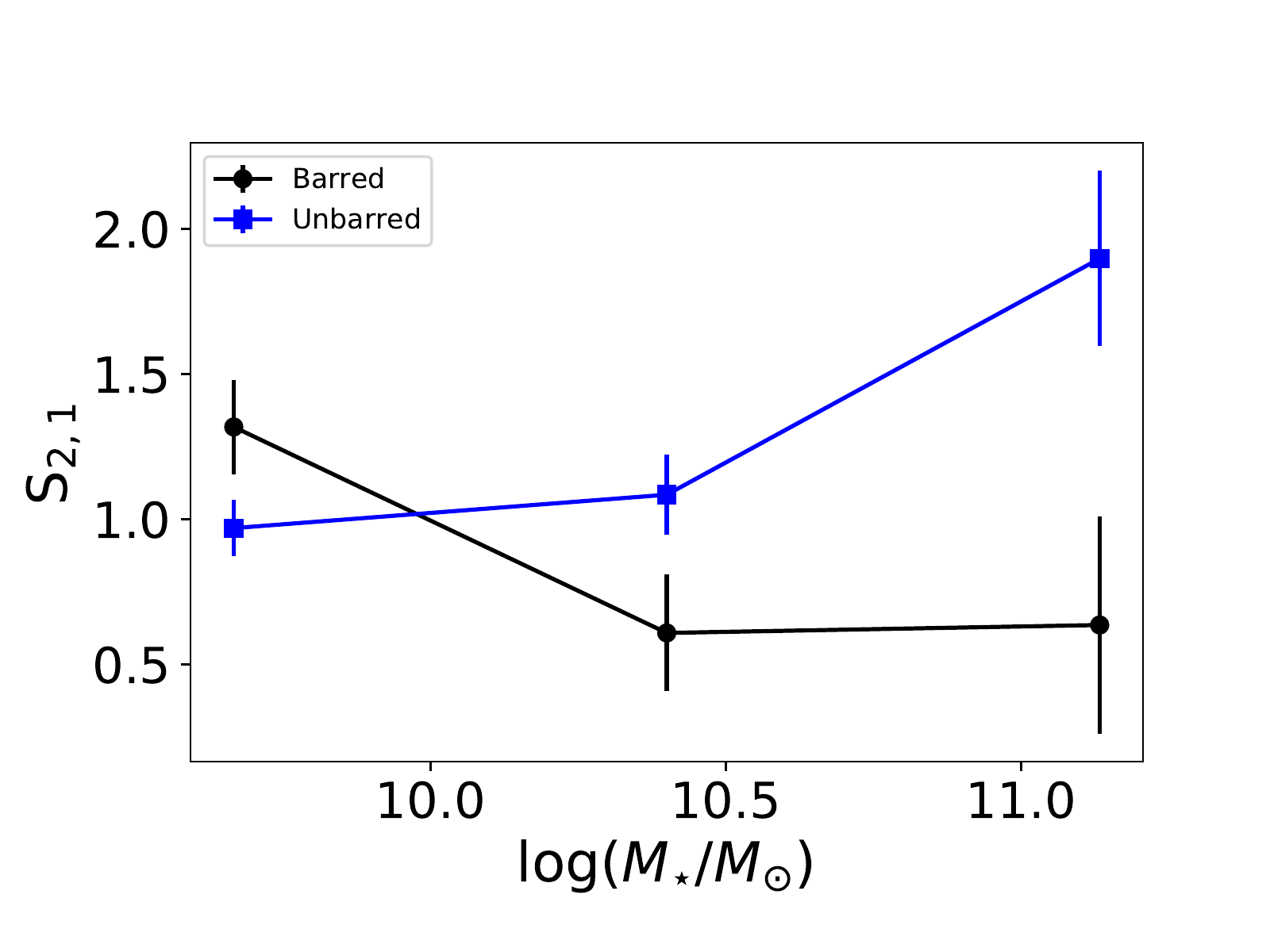} &
\includegraphics[width=0.33\textwidth]{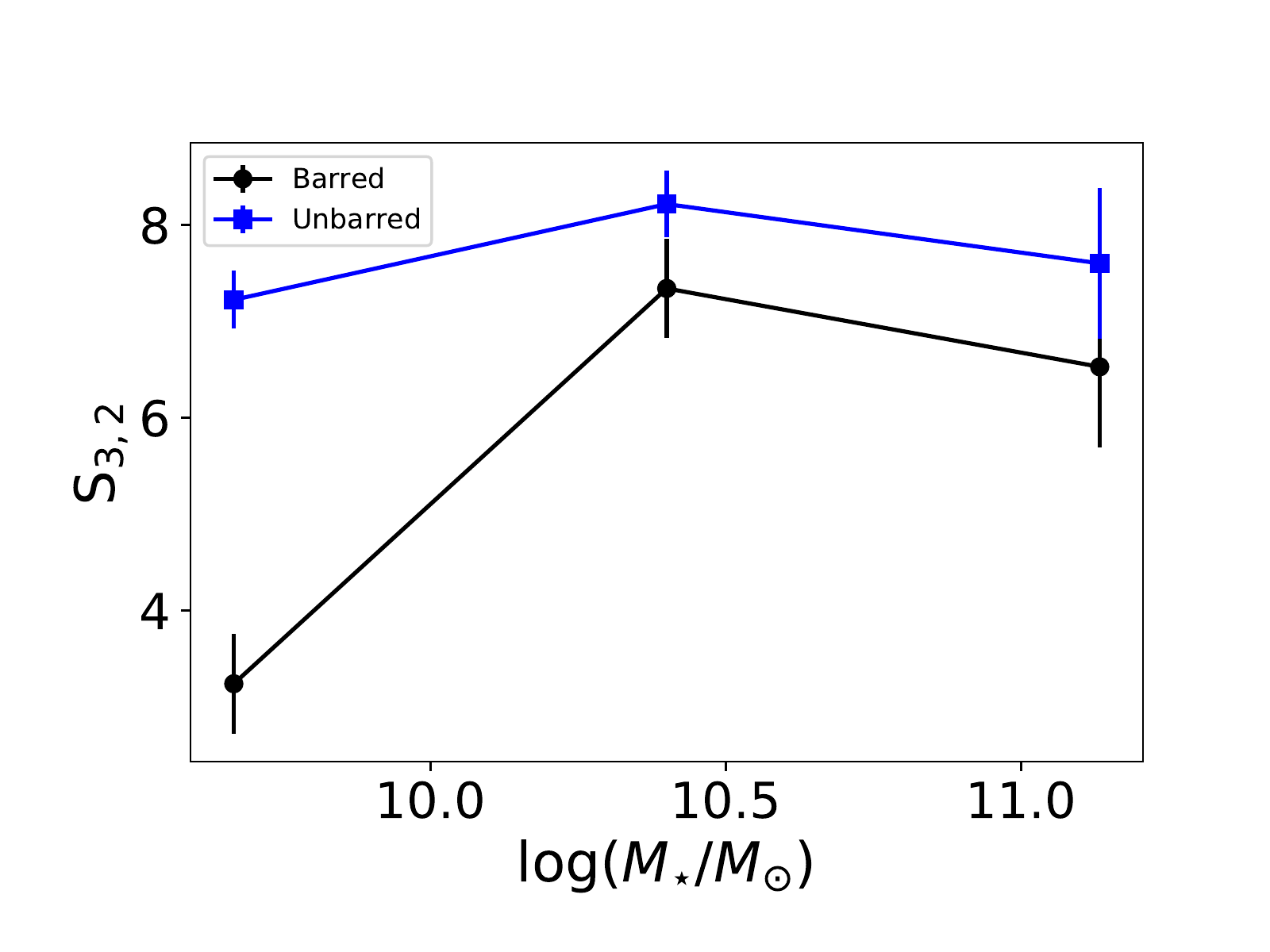} &
\includegraphics[width=0.33\textwidth]{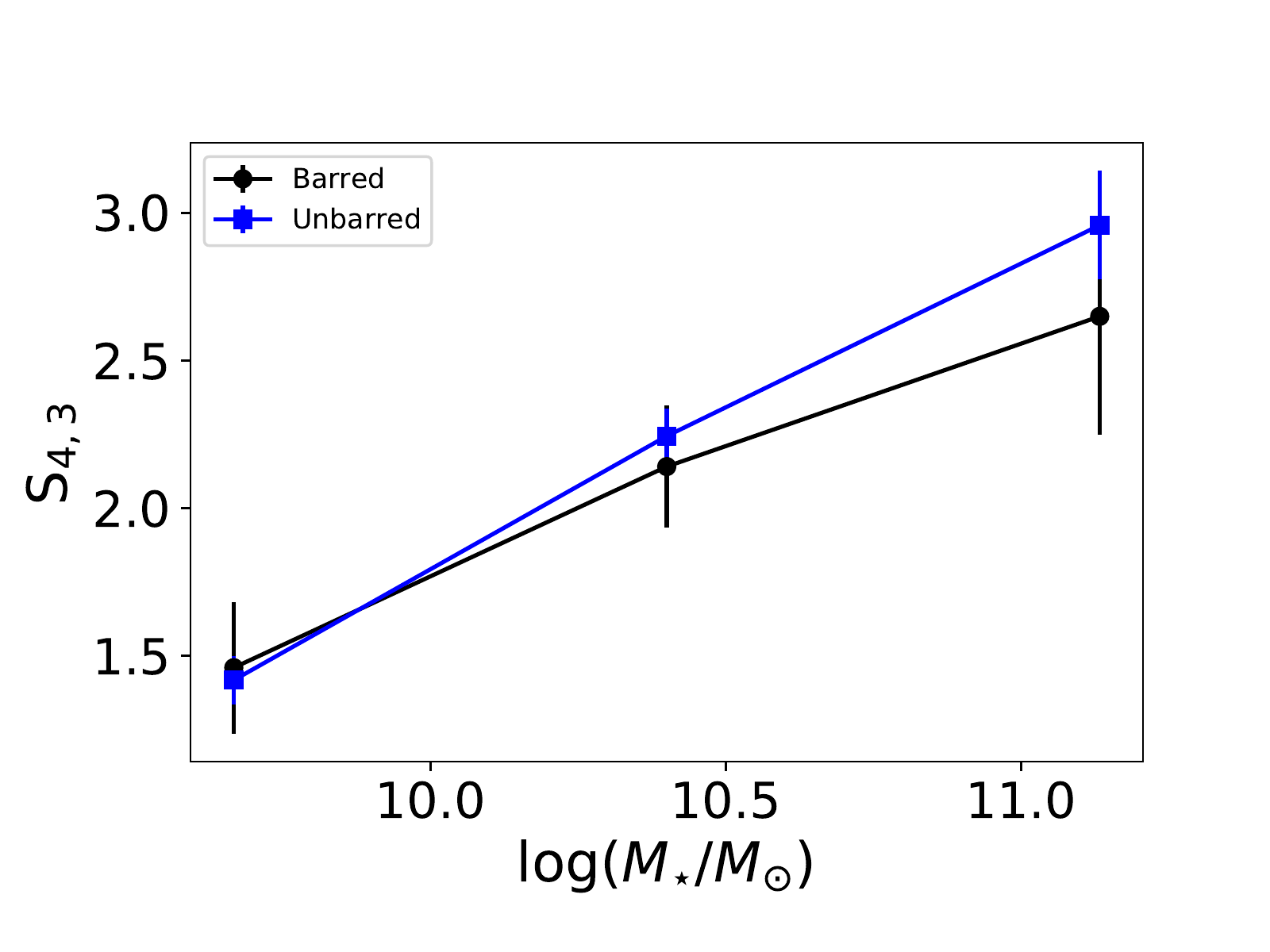} \\
\end{tabular}
\caption{Quenching indexes and SFR for barred and unbarred cluster galaxies as a function of the stellar mass.}
\label{fig:SFR_ratio}
\end{figure*}

Quenching indexes were calculated for barred and unbarred galaxies for three different stellar mass bins. Separating into  mass bins allows us to disentangle downsizing effects, by which more massive galaxies are expected to have the bulk of their stellar populations formed at earlier epochs, hence displaying on average higher quenching index values. We present these results in \autoref{fig:SFR_ratio}. 

We have analyzed three possible combinations for the quenching indexes: $S_{4,3}$, $S_{3,2}$, and $S_{2,1}$ comparing barred and unbarred to look for possible signs of enhancement/quenching or differences provoked by the presence of the bar. As expected, for the older index, $S_{4,3}$, there is no difference when barred and unbarred galaxies are considered, at any mass: this, in fact, represents the star formation mode at the epoch of the galaxy assembly when a bar was very likely not formed yet, and no differences are hence expected. A small difference is found in the most massive bin, with barred galaxies presenting slightly lower index, even though the average values are within the dispersions. This might be due to the fact that stellar bars redistribute stellar populations and might bring in the innermost regions of the galaxies younger stars that were initially located outside the bar region. The fact that this is only observed in the most massive bin is probably a consequence of the fact that stronger bars are preferentially observed in more massive galaxies.

When we go to more recent epochs analyzing the $S_{3,2}$ index, we do instead find differences at all masses, in that unbarred galaxies show higher values pointing at a more efficient quenching as respect to barred ones, this difference being a function of the stellar mass: low mass galaxies show the largest difference (a factor of more than 2), while in the two high mass bins differences are only marginal (i.e. within the uncertainties), but still significant. 

The behavior for the most recent index, $S_{2,1}$ is not as linear: at low masses, SFR is more inhibited in barred galaxies, and the trend is reversed in the two highest mass bins, with barred galaxies showing SFR enhancement, as it is expected by the centrally rejuvenation by the presence of the bar. The case of low mass galaxies is different as they are more easily perturbed by external factors due to their shallower gravitational potentials: barred galaxies at the lowest mass, are more easily induced to form stars because of the bar itself. Hence, before any morphological transformation occurs, they are subject to a relatively efficient star formation episode as compared to unbarred galaxies (see the plot for $S_{3,2}$). As a consequence, their gas reservoir is already partially depleted compared to the one of unbarred galaxies, which are still able to form stars at a higher rate in the most recent epoch as accounted by $S_{2,1}$.

\section{Summary and Conclusions}
\label{sec:conclusions}

We used a semi-automatic bar detection to analyze the V-band images of 32 clusters drown from OmegaWINGS survey. The adopted algorithm is a relatively standard one, based on the detection of changes in the ellipticity radial profile at constant position angle. Our sample was restricted to cluster members with ellipticity $\epsilon <$ 0.5, surface brightness in V-band SB $<$ 21.5, and $M_{type}$ $>$ -5. To validate our procedure, we visually inspected the images and the isophotes of our sample, and found errors/misclassification in less than 10\% of the galaxies. 
The final number of disk galaxies used for this work is 3,456 of which 906 were identified as barred (26$\%$). This fraction is close to the values found in a number of other studies, both in cluster and in the field \citep{Barazza2009, Marinova2009, Masters2010, Alonso2013, Vera2016, Cervantes2015, Yoon2019}. 

Having identified barred galaxies in our sample, we performed a detailed analysis to understand how the bar fraction depends on the intrinsic galaxy properties (total stellar mass (log($M_{\star}/M_{\odot}$), color (B-V) and the morphological type ($M_{type}$)), the global properties of the galaxy clusters (X-ray luminosity) and the cluster environment (the projected clustercentric distance ($(r_p/r_{virial})_{cl}$) and the nearest neighbor galaxy($r_p/r_{virial}$)$_{ng}$). 

We found that the bar fraction increases toward massive, bluer galaxies and as we go from early- to late-type galaxies. The dependence on color vanishes once we fix the morphological type, indicating that the dependence of $f_{bar}$ on (B-V) is mostly driven by morphology for cluster galaxies.

The bar fraction shows a mild declining trend as a function of the X-ray luminosity, suggesting at first that more massive clusters are a hostile environment for the survival of bars in galaxies. A further analysis shows that this trend is mimicked by a diminishing fraction of late-type galaxies, which are the objects where bars are most likely to form, hence providing a natural and straightforward interpretation to this dependence.  Now, it should be noted that the more massive clusters have a larger virial radius, and hence the coverage in their outskirts is at least partially incomplete. If we restrict our analysis to a homogeneous coverage within 0.7 $R_{vir}$, the trend weakens.

 Similarly, a morphological type--bar connection holds when analyzing the barred fraction as a function of the clustercentric distance: the higher fraction of barred galaxies is always found among the latest morphological type, at any projected distance, and almost monotonically decreases going from late- to early-type, independently of the location within the cluster. This effect is clearly a reflection of the morphology-density relation \citep{dressler80} and of the infall time on galaxies which are in the process of going through a morphological transformation which will eventually lead them to be dynamically hot systems with an early-type morphology. This interpretation is further reinforced if looking at where barred galaxies are preferably found in the phase--space diagram: the highest fraction is indeed concentrated at distances further than one virial radius and with relatively high velocities, making them the best candidate for recently infallen objects \citep{Rhee2017, Jaffe2019}.

In order to investigate the influence of bar fraction in galaxy-galaxy interactions, we computed the bar fraction as a function of the distance to the nearest neighbor galaxy normalized by the virial radius of the neighbor galaxy. A fairly constant $f_{bar}$ for galaxies with a separation larger than one viral radius was found, while for close pairs, the fraction of barred galaxies presents a dramatic decline. Hence, close gravitational encounters between galaxies are able to perturb galaxies to a point that bars are destroyed \citep{Lee2012, Lin2014}, even in the cluster environment \citep{Yoon2019}, although the co-dependence with clustercentric projected distance is difficult to disentangle and an underlying dependence with the local density could mimic the dependence with the distance to the nearest neighbor.

Bars are known to be particularly efficient at redistributing matter and angular momentum between different components, including gas that, given its collisional nature, can flow inward, increasing the available molecular gas in the central regions, hence potentially increasing the star formation activity. On the other side, galaxies in clusters are subject to several mechanisms that are very efficient in perturbing and even removing gas both in their disks and in their halo as well, something which will eventually lead them to be unable to sustain detectable levels of star formation activity. It is hence very interesting to investigate the result of the combined interaction between an internal mechanism such as a bar, and the variety of external physical phenomena (harassment, ram pressure, tidal interactions, etc.) which galaxies are subject to in clusters.

We tackle this aspect by analyzing the central specific star formation rate and the luminosity weighted stellar age as a function of morphology, clustercentric distance and the presence or absence of bars. We identify two maxima in the bar fraction: one in a region in which galaxies are late-types, characterized by low star formation activity and old stellar populations, located in the outskirts of the clusters. These barred galaxies with quenched central activity and old ages are the result of an enhanced star formation activity in the past as a result of the formation and grow of their stellar bars as opposed to their unbarred counterparts. The second maximum of the bar fraction occurs for early-type galaxies with high central star formation activity and young stellar ages mostly located within the virial radii of the clusters, hence likely to be galaxies of recent infall. A plausible scenario for these kind of galaxies is one in which already centrally quenched barred late-type galaxies are in the process of morphological transformation by the cluster environment, that not only perturb their stellar component, but also their gas, feeding the region swept by the bar that in turns triggers the star formation activity as detected in the galaxies of our sample.

Our results indicate that the likelihood of a galaxy hosting a bar in the cluster environment is strongly dependent on morphology. As galaxies experience a morphological transformation once they fall into the gravitational well of the clusters, the bar itself can be weakened or even destroyed once the dynamically cold disk is perturbed. This would explain why the bar fraction in our sample is maximum in the outskirts of the clusters and for galaxies that are just now falling into these structures. Once the gravitational perturbation is enough to produce a morphological transformation, the dissolution of the bar will proceed, not before inducing central star formation activity due to the bar acting on the perturbed gas that is now available in this region, as a result of the interplay between internal and environmental secular evolution.

\begin{acknowledgments}
B. Cervantes Sodi and Amira A. Tawfeek thank Osbaldo S\'anchez Garc\'ia for useful comments and discussions during the early phases of this project. J.F., Amira A. Tawfeek and D.P.M acknowledge financial support from the UNAM-DGAPA-PAPIIT IN111620 grant, M\'exico. B. Cervantes Sodi and Amira A. Tawfeek acknowledge financial support through PAPIIT project IA103520, from DGAPA-UNAM, M\'exico.
\end{acknowledgments}

\bibliography{tawfeek_22}{}

\begin{thebibliography}{}
\expandafter\ifx\csname natexlab\endcsname\relax\def\natexlab#1{#1}\fi
\providecommand{\url}[1]{\href{#1}{#1}}
\providecommand{\dodoi}[1]{doi:~\href{http://doi.org/#1}{\nolinkurl{#1}}}
\providecommand{\doeprint}[1]{\href{http://ascl.net/#1}{\nolinkurl{http://ascl.net/#1}}}
\providecommand{\doarXiv}[1]{\href{https://arxiv.org/abs/#1}{\nolinkurl{https://arxiv.org/abs/#1}}}

\bibitem[{{Aguerri} \& {Gonz{\'a}lez-Garc{\'\i}a}(2009)}]{Aguerri2009a}
{Aguerri}, J.~A.~L., \& {Gonz{\'a}lez-Garc{\'\i}a}, A.~C. 2009, \aap, 494, 891,
  \dodoi{10.1051/0004-6361:200810339}

\bibitem[{{Aguerri} {et~al.}(2009){Aguerri}, {M{\'e}ndez-Abreu}, \&
  {Corsini}}]{Aguerri2009}
{Aguerri}, J.~A.~L., {M{\'e}ndez-Abreu}, J., \& {Corsini}, E.~M. 2009, \aap,
  495, 491, \dodoi{10.1051/0004-6361:200810931}

\bibitem[{{Alonso} {et~al.}(2013){Alonso}, {Coldwell}, \&
  {Lambas}}]{Alonso2013}
{Alonso}, M.~S., {Coldwell}, G., \& {Lambas}, D.~G. 2013, \aap, 549, A141,
  \dodoi{10.1051/0004-6361/201220117}

\bibitem[{{Ann} {et~al.}(2015){Ann}, {Seo}, \& {Ha}}]{Ann2015}
{Ann}, H.~B., {Seo}, M., \& {Ha}, D.~K. 2015, \apjs, 217, 27,
  \dodoi{10.1088/0067-0049/217/2/27}

\bibitem[{{Athanassoula}(2003)}]{Athanassoula2003}
{Athanassoula}, E. 2003, in Revista Mexicana de Astronomia y Astrofisica
  Conference Series, Vol.~17, Revista Mexicana de Astronomia y Astrofisica
  Conference Series, ed. V.~{Avila-Reese}, C.~{Firmani}, C.~S. {Frenk}, \&
  C.~{Allen}, 28--29

\bibitem[{{Athanassoula} {et~al.}(1983){Athanassoula}, {Bienayme}, {Martinet},
  \& {Pfenniger}}]{Athanassoula1983}
{Athanassoula}, E., {Bienayme}, O., {Martinet}, L., \& {Pfenniger}, D. 1983,
  \aap, 127, 349

\bibitem[{{Athanassoula} \& {Misiriotis}(2002)}]{Athanassoula2002}
{Athanassoula}, E., \& {Misiriotis}, A. 2002, \mnras, 330, 35,
  \dodoi{10.1046/j.1365-8711.2002.05028.x}

\bibitem[{{Barazza} {et~al.}(2009){Barazza}, {Jablonka}, {Desai}, {Jogee},
  {Arag{\'o}n-Salamanca}, {De Lucia}, {Saglia}, {Halliday}, {Poggianti},
  {Dalcanton}, {Rudnick}, {Milvang-Jensen}, {Noll}, {Simard}, {Clowe},
  {Pell{\'o}}, {White}, \& {Zaritsky}}]{Barazza2009}
{Barazza}, F.~D., {Jablonka}, P., {Desai}, V., {et~al.} 2009, \aap, 497, 713,
  \dodoi{10.1051/0004-6361/200810352}

\bibitem[{{Barway} \& {Saha}(2020)}]{Barway2020}
{Barway}, S., \& {Saha}, K. 2020, \mnras, 495, 4548,
  \dodoi{10.1093/mnras/staa1387}

\bibitem[{{Buta} {et~al.}(2010){Buta}, {Sheth}, {Regan}, {Hinz}, {Gil de Paz},
  {Men{\'e}ndez-Delmestre}, {Munoz-Mateos}, {Seibert}, {Laurikainen}, {Salo},
  {Gadotti}, {Athanassoula}, {Bosma}, {Knapen}, {Ho}, {Madore}, {Elmegreen},
  {Masters}, {Comer{\'o}n}, {Aravena}, \& {Kim}}]{Buta2010}
{Buta}, R.~J., {Sheth}, K., {Regan}, M., {et~al.} 2010, \apjs, 190, 147,
  \dodoi{10.1088/0067-0049/190/1/147}

\bibitem[{{Butcher} \& {Oemler}(1984)}]{butcher84}
{Butcher}, H., \& {Oemler}, A., J. 1984, \apj, 285, 426, \dodoi{10.1086/162519}

\bibitem[{{Capaccioli} \& {Schipani}(2011)}]{Capaccioli2011}
{Capaccioli}, M., \& {Schipani}, P. 2011, The Messenger, 146, 2

\bibitem[{{Cava} {et~al.}(2009){Cava}, {Bettoni}, {Poggianti}, {Couch},
  {Moles}, {Varela}, {Biviano}, {D'Onofrio}, {Dressler}, {Fasano}, {Fritz},
  {Kj{\ae}rgaard}, {Ramella}, \& {Valentinuzzi}}]{Cava2009}
{Cava}, A., {Bettoni}, D., {Poggianti}, B.~M., {et~al.} 2009, \aap, 495, 707,
  \dodoi{10.1051/0004-6361:200810997}

\bibitem[{{Cervantes Sodi} {et~al.}(2015){Cervantes Sodi}, {Li}, \&
  {Park}}]{Cervantes2015}
{Cervantes Sodi}, B., {Li}, C., \& {Park}, C. 2015, \apj, 807, 111,
  \dodoi{10.1088/0004-637X/807/1/111}

\bibitem[{{Cervantes-Sodi} {et~al.}(2013){Cervantes-Sodi}, {Li}, {Park}, \&
  {Wang}}]{Cervantes2013}
{Cervantes-Sodi}, B., {Li}, C., {Park}, C., \& {Wang}, L. 2013, \apj, 775, 19,
  \dodoi{10.1088/0004-637X/775/1/19}

\bibitem[{{Chabrier}(2003)}]{chabrier03}
{Chabrier}, G. 2003, \pasp, 115, 763, \dodoi{10.1086/376392}

\bibitem[{{Combes}(2003)}]{Combes2003}
{Combes}, F. 2003, in SF2A-2003: Semaine de l'Astrophysique Francaise, ed.
  F.~{Combes}, D.~{Barret}, T.~{Contini}, \& L.~{Pagani}, 243.
\newblock \doarXiv{astro-ph/0308022}

\bibitem[{{Contopoulos} \& {Grosbol}(1989)}]{Contopoulos1989}
{Contopoulos}, G., \& {Grosbol}, P. 1989, \aapr, 1, 261,
  \dodoi{10.1007/BF00873080}

\bibitem[{{Dalcanton} {et~al.}(2004){Dalcanton}, {Yoachim}, \&
  {Bernstein}}]{Dalcanton2004}
{Dalcanton}, J.~J., {Yoachim}, P., \& {Bernstein}, R.~A. 2004, \apj, 608, 189,
  \dodoi{10.1086/386358}

\bibitem[{{Dressler}(1980)}]{dressler80}
{Dressler}, A. 1980, \apj, 236, 351, \dodoi{10.1086/157753}

\bibitem[{{Ellison} {et~al.}(2011){Ellison}, {Patton}, {Mendel}, \&
  {Scudder}}]{Ellison2011}
{Ellison}, S.~L., {Patton}, D.~R., {Mendel}, J.~T., \& {Scudder}, J.~M. 2011,
  \mnras, 418, 2043, \dodoi{10.1111/j.1365-2966.2011.19624.x}

\bibitem[{{Elmegreen} {et~al.}(1990){Elmegreen}, {Elmegreen}, \&
  {Bellin}}]{Elmegreen1990}
{Elmegreen}, D.~M., {Elmegreen}, B.~G., \& {Bellin}, A.~D. 1990, \apj, 364,
  415, \dodoi{10.1086/169424}

\bibitem[{{Eskridge} {et~al.}(2000){Eskridge}, {Frogel}, {Pogge}, {Quillen},
  {Davies}, {DePoy}, {Houdashelt}, {Kuchinski}, {Ram{\'\i}rez}, {Sellgren},
  {Terndrup}, \& {Tiede}}]{Eskridge2000}
{Eskridge}, P.~B., {Frogel}, J.~A., {Pogge}, R.~W., {et~al.} 2000, \aj, 119,
  536, \dodoi{10.1086/301203}

\bibitem[{{Fasano} {et~al.}(2006){Fasano}, {Marmo}, {Varela}, {D'Onofrio},
  {Poggianti}, {Moles}, {Pignatelli}, {Bettoni}, {Kj{\ae}rgaard}, {Rizzi},
  {Couch}, \& {Dressler}}]{Fasano2006}
{Fasano}, G., {Marmo}, C., {Varela}, J., {et~al.} 2006, \aap, 445, 805,
  \dodoi{10.1051/0004-6361:20053816}

\bibitem[{{Fasano} {et~al.}(2012){Fasano}, {Vanzella}, {Dressler}, {Poggianti},
  {Moles}, {Bettoni}, {Valentinuzzi}, {Moretti}, {D'Onofrio}, {Varela},
  {Couch}, {Kj{\ae}rgaard}, {Fritz}, {Omizzolo}, \& {Cava}}]{Fasano2012}
{Fasano}, G., {Vanzella}, E., {Dressler}, A., {et~al.} 2012, \mnras, 420, 926,
  \dodoi{10.1111/j.1365-2966.2011.19798.x}

\bibitem[{{Fasano} {et~al.}(2015){Fasano}, {Poggianti}, {Bettoni}, {D'Onofrio},
  {Dressler}, {Vulcani}, {Moretti}, {Gullieuszik}, {Fritz}, {Omizzolo}, {Cava},
  {Couch}, {Ramella}, \& {Biviano}}]{fasano15}
{Fasano}, G., {Poggianti}, B.~M., {Bettoni}, D., {et~al.} 2015, \mnras, 449,
  3927, \dodoi{10.1093/mnras/stv500}

\bibitem[{{Finn} {et~al.}(2005){Finn}, {Zaritsky}, {McCarthy}, {Poggianti},
  {Rudnick}, {Halliday}, {Milvang-Jensen}, {Pell{\'o}}, \& {Simard}}]{Finn2005}
{Finn}, R.~A., {Zaritsky}, D., {McCarthy}, Donald~W., J., {et~al.} 2005, \apj,
  630, 206, \dodoi{10.1086/431642}

\bibitem[{{Fraser-McKelvie} {et~al.}(2020){Fraser-McKelvie}, {Merrifield},
  {Arag{\'o}n-Salamanca}, {Peterken}, {Kraljic}, {Masters}, {Stark},
  {Fragkoudi}, {Smethurst}, {Boardman}, {Drory}, \& {Lane}}]{Fraser2020}
{Fraser-McKelvie}, A., {Merrifield}, M., {Arag{\'o}n-Salamanca}, A., {et~al.}
  2020, \mnras, 499, 1116, \dodoi{10.1093/mnras/staa2866}

\bibitem[{{Friedli} \& {Benz}(1995)}]{Friedli1995}
{Friedli}, D., \& {Benz}, W. 1995, \aap, 301, 649

\bibitem[{{Fritz} {et~al.}(2007){Fritz}, {Poggianti}, {Bettoni}, {Cava},
  {Couch}, {D'Onofrio}, {Dressler}, {Fasano}, {Kj{\ae}rgaard}, {Moles}, \&
  {Varela}}]{fritz07}
{Fritz}, J., {Poggianti}, B.~M., {Bettoni}, D., {et~al.} 2007, \aap, 470, 137,
  \dodoi{10.1051/0004-6361:20077097}

\bibitem[{{Fritz} {et~al.}(2011){Fritz}, {Poggianti}, {Cava}, {Valentinuzzi},
  {Moretti}, {Bettoni}, {Bressan}, {Couch}, {D'Onofrio}, {Dressler}, {Fasano},
  {Kj{\ae}rgaard}, {Moles}, {Omizzolo}, \& {Varela}}]{fritz11}
{Fritz}, J., {Poggianti}, B.~M., {Cava}, A., {et~al.} 2011, \aap, 526, A45,
  \dodoi{10.1051/0004-6361/201015214}

\bibitem[{{Fritz} {et~al.}(2017){Fritz}, {Moretti}, {Gullieuszik}, {Poggianti},
  {Bruzual}, {Vulcani}, {Nicastro}, {Jaff{\'e}}, {Cervantes Sodi}, {Bettoni},
  {Biviano}, {Fasano}, {Charlot}, {Bellhouse}, \& {Hau}}]{fritz17}
{Fritz}, J., {Moretti}, A., {Gullieuszik}, M., {et~al.} 2017, \apj, 848, 132,
  \dodoi{10.3847/1538-4357/aa8f51}

\bibitem[{{Gavazzi} {et~al.}(2015){Gavazzi}, {Consolandi}, {Dotti}, {Fanali},
  {Fossati}, {Fumagalli}, {Viscardi}, {Savorgnan}, {Boselli}, {Guti{\'e}rrez},
  {Hern{\'a}ndez Toledo}, {Giovanelli}, \& {Haynes}}]{Gavazzi2015}
{Gavazzi}, G., {Consolandi}, G., {Dotti}, M., {et~al.} 2015, \aap, 580, A116,
  \dodoi{10.1051/0004-6361/201425351}

\bibitem[{{George} {et~al.}(2019){George}, {Joseph}, {Mondal}, {Subramanian},
  {Subramaniam}, \& {Paul}}]{George2019}
{George}, K., {Joseph}, P., {Mondal}, C., {et~al.} 2019, \aap, 621, L4,
  \dodoi{10.1051/0004-6361/201834500}

\bibitem[{{George} {et~al.}(2020){George}, {Joseph}, {Mondal}, {Subramanian},
  {Subramaniam}, \& {Paul}}]{George2020}
---. 2020, \aap, 644, A79, \dodoi{10.1051/0004-6361/202038810}

\bibitem[{{Gullieuszik} {et~al.}(2015){Gullieuszik}, {Poggianti}, {Fasano},
  {Zaggia}, {Paccagnella}, {Moretti}, {Bettoni}, {D'Onofrio}, {Couch},
  {Vulcani}, {Fritz}, {Omizzolo}, {Baruffolo}, {Schipani}, {Capaccioli}, \&
  {Varela}}]{Gullieuszik2015}
{Gullieuszik}, M., {Poggianti}, B., {Fasano}, G., {et~al.} 2015, \aap, 581,
  A41, \dodoi{10.1051/0004-6361/201526061}

\bibitem[{{Ho} {et~al.}(1997){Ho}, {Filippenko}, \& {Sargent}}]{Ho1997}
{Ho}, L.~C., {Filippenko}, A.~V., \& {Sargent}, W. L.~W. 1997, \apj, 487, 591,
  \dodoi{10.1086/304643}

\bibitem[{{Hunt} {et~al.}(2008){Hunt}, {Combes}, {Garc{\'\i}a-Burillo},
  {Schinnerer}, {Krips}, {Baker}, {Boone}, {Eckart}, {L{\'e}on}, {Neri}, \&
  {Tacconi}}]{Hunt2008}
{Hunt}, L.~K., {Combes}, F., {Garc{\'\i}a-Burillo}, S., {et~al.} 2008, \aap,
  482, 133, \dodoi{10.1051/0004-6361:20078874}

\bibitem[{{Jaff{\'e}} {et~al.}(2019){Jaff{\'e}}, {Poggianti}, {Moretti},
  {Gullieuszik}, {Smith}, {Vulcani}, {Fasano}, {Fritz}, {Tonnesen}, {Bettoni},
  {Hau}, {Biviano}, {Bellhouse}, \& {McGee}}]{Jaffe2019}
{Jaff{\'e}}, Y.~L., {Poggianti}, B.~M., {Moretti}, A., {et~al.} 2019, \mnras,
  482, 3454, \dodoi{10.1093/mnras/sty2774}

\bibitem[{{James} \& {Percival}(2015)}]{James2015}
{James}, P.~A., \& {Percival}, S.~M. 2015, \mnras, 450, 3503,
  \dodoi{10.1093/mnras/stv846}

\bibitem[{{James} \& {Percival}(2018)}]{James2018}
---. 2018, \mnras, 474, 3101, \dodoi{10.1093/mnras/stx2990}

\bibitem[{{Jedrzejewski}(1987)}]{Jedrzejewski1987}
{Jedrzejewski}, R.~I. 1987, \mnras, 226, 747, \dodoi{10.1093/mnras/226.4.747}

\bibitem[{{Jogee} {et~al.}(2005){Jogee}, {Scoville}, \& {Kenney}}]{Jogee2005}
{Jogee}, S., {Scoville}, N., \& {Kenney}, J. D.~P. 2005, \apj, 630, 837,
  \dodoi{10.1086/432106}

\bibitem[{{Jogee} {et~al.}(2004){Jogee}, {Barazza}, {Rix}, {Davies}, {Heyer},
  {Barden}, {Beckwith}, {Bell}, {Borch}, {Caldwell}, {Conselice},
  {H{\"a}ussler}, {Heymans}, {Jahnke}, {Knapen}, {Laine}, {Lubell}, {Mobasher},
  {McIntosh}, {Meisenheimer}, {Peng}, {Ravindranath}, {Sanchez}, {Shlosman},
  {Somerville}, {Wisotzki}, \& {Wolf}}]{Jogee2004}
{Jogee}, S., {Barazza}, F.~D., {Rix}, H.~W., {et~al.} 2004, {Evolution and
  Impact of Bars over the last nine Gyr: Early Results from GEMS}, Vol. 319,
  291, \dodoi{10.1007/978-1-4020-2862-5_26}

\bibitem[{{Kim} {et~al.}(2017){Kim}, {Hwang}, {Chung}, {Lee}, {Park},
  {Cervantes Sodi}, \& {Kim}}]{Kim2017}
{Kim}, E., {Hwang}, H.~S., {Chung}, H., {et~al.} 2017, \apj, 845, 93,
  \dodoi{10.3847/1538-4357/aa80db}

\bibitem[{{Kim} {et~al.}(2020){Kim}, {Choi}, \& {Kim}}]{Kim2020}
{Kim}, M., {Choi}, Y.-Y., \& {Kim}, S.~S. 2020, arXiv e-prints,
  arXiv:2008.13743.
\newblock \doarXiv{2008.13743}

\bibitem[{{Kormendy} \& {Kennicutt}(2004)}]{Kormendy2004}
{Kormendy}, J., \& {Kennicutt}, Robert~C., J. 2004, \araa, 42, 603,
  \dodoi{10.1146/annurev.astro.42.053102.134024}

\bibitem[{{Laine} {et~al.}(2002){Laine}, {Shlosman}, {Knapen}, \&
  {Peletier}}]{Laine2002}
{Laine}, S., {Shlosman}, I., {Knapen}, J.~H., \& {Peletier}, R.~F. 2002, \apj,
  567, 97, \dodoi{10.1086/323964}

\bibitem[{{Lansbury} {et~al.}(2014){Lansbury}, {Lucey}, \&
  {Smith}}]{Lansbury2014}
{Lansbury}, G.~B., {Lucey}, J.~R., \& {Smith}, R.~J. 2014, \mnras, 439, 1749,
  \dodoi{10.1093/mnras/stu049}

\bibitem[{{Lee} {et~al.}(2012){Lee}, {Park}, {Lee}, \& {Choi}}]{Lee2012}
{Lee}, G.-H., {Park}, C., {Lee}, M.~G., \& {Choi}, Y.-Y. 2012, \apj, 745, 125,
  \dodoi{10.1088/0004-637X/745/2/125}

\bibitem[{{Lee} {et~al.}(2019){Lee}, {Ann}, \& {Park}}]{Lee2019}
{Lee}, Y.~H., {Ann}, H.~B., \& {Park}, M.-G. 2019, \apj, 872, 97,
  \dodoi{10.3847/1538-4357/ab0024}

\bibitem[{{Lin} {et~al.}(2014){Lin}, {Cervantes Sodi}, {Li}, {Wang}, \&
  {Wang}}]{Lin2014}
{Lin}, Y., {Cervantes Sodi}, B., {Li}, C., {Wang}, L., \& {Wang}, E. 2014,
  \apj, 796, 98, \dodoi{10.1088/0004-637X/796/2/98}

\bibitem[{{{\L}okas}(2020)}]{Lokas2020}
{{\L}okas}, E.~L. 2020, \aap, 642, L12, \dodoi{10.1051/0004-6361/202039425}

\bibitem[{{{\L}okas} {et~al.}(2016){{\L}okas}, {Ebrov{\'a}}, {del Pino},
  {Sybilska}, {Athanassoula}, {Semczuk}, {Gajda}, \& {Fouquet}}]{Lokas2016}
{{\L}okas}, E.~L., {Ebrov{\'a}}, I., {del Pino}, A., {et~al.} 2016, \apj, 826,
  227, \dodoi{10.3847/0004-637X/826/2/227}

\bibitem[{{Lynden-Bell} \& {Kalnajs}(1972)}]{Lynden1972}
{Lynden-Bell}, D., \& {Kalnajs}, A.~J. 1972, \mnras, 157, 1,
  \dodoi{10.1093/mnras/157.1.1}

\bibitem[{{Marinova} \& {Jogee}(2007)}]{Marinova2007}
{Marinova}, I., \& {Jogee}, S. 2007, \apj, 659, 1176, \dodoi{10.1086/512355}

\bibitem[{{Marinova} {et~al.}(2009){Marinova}, {Jogee}, {Heiderman}, {Barazza},
  {Gray}, {Barden}, {Wolf}, {Peng}, {Bacon}, {Balogh}, {Bell}, {B{\"o}hm},
  {Caldwell}, {H{\"a}u{\ss}ler}, {Heymans}, {Jahnke}, {van Kampen}, {Lane},
  {McIntosh}, {Meisenheimer}, {S{\'a}nchez}, {Somerville}, {Taylor},
  {Wisotzki}, \& {Zheng}}]{Marinova2009}
{Marinova}, I., {Jogee}, S., {Heiderman}, A., {et~al.} 2009, \apj, 698, 1639,
  \dodoi{10.1088/0004-637X/698/2/1639}

\bibitem[{{Marinova} {et~al.}(2012){Marinova}, {Jogee}, {Weinzirl}, {Erwin},
  {Trentham}, {Ferguson}, {Hammer}, {den Brok}, {Graham}, {Carter}, {Balcells},
  {Goudfrooij}, {Guzm{\'a}n}, {Hoyos}, {Mobasher}, {Mouhcine}, {Peletier},
  {Peng}, \& {Verdoes Kleijn}}]{Marinova2012}
{Marinova}, I., {Jogee}, S., {Weinzirl}, T., {et~al.} 2012, \apj, 746, 136,
  \dodoi{10.1088/0004-637X/746/2/136}

\bibitem[{{Martinez-Valpuesta} {et~al.}(2016){Martinez-Valpuesta}, {Aguerri},
  \& {Gonz{\'a}lez-Garc{\'\i}a}}]{Martinez2016}
{Martinez-Valpuesta}, I., {Aguerri}, J., \& {Gonz{\'a}lez-Garc{\'\i}a}, C.
  2016, Galaxies, 4, 7, \dodoi{10.3390/galaxies4020007}

\bibitem[{{Martini} {et~al.}(2001){Martini}, {Mulchaey}, {Pogge}, \&
  {Regan}}]{Martini2001}
{Martini}, P., {Mulchaey}, J.~S., {Pogge}, R.~W., \& {Regan}, M.~W. 2001, in
  American Astronomical Society Meeting Abstracts, Vol. 198, American
  Astronomical Society Meeting Abstracts \#198, 85.03

\bibitem[{{Masters} {et~al.}(2010){Masters}, {Mosleh}, {Romer}, {Nichol},
  {Bamford}, {Schawinski}, {Lintott}, {Andreescu}, {Campbell}, {Crowcroft},
  {Doyle}, {Edmondson}, {Murray}, {Raddick}, {Slosar}, {Szalay}, \&
  {Vandenberg}}]{Masters2010}
{Masters}, K.~L., {Mosleh}, M., {Romer}, A.~K., {et~al.} 2010, \mnras, 405,
  783, \dodoi{10.1111/j.1365-2966.2010.16503.x}

\bibitem[{{Masters} {et~al.}(2011){Masters}, {Maraston}, {Nichol}, {Thomas},
  {Beifiori}, {Bundy}, {Edmondson}, {Higgs}, {Leauthaud}, {Mandelbaum},
  {Pforr}, {Ross}, {Ross}, {Schneider}, {Skibba}, {Tinker}, {Tojeiro}, {Wake},
  {Brinkmann}, \& {Weaver}}]{Masters2011}
{Masters}, K.~L., {Maraston}, C., {Nichol}, R.~C., {et~al.} 2011, \mnras, 418,
  1055, \dodoi{10.1111/j.1365-2966.2011.19557.x}

\bibitem[{{Masters} {et~al.}(2012){Masters}, {Nichol}, {Haynes}, {Keel},
  {Lintott}, {Simmons}, {Skibba}, {Bamford}, {Giovanelli}, \&
  {Schawinski}}]{Masters2012}
{Masters}, K.~L., {Nichol}, R.~C., {Haynes}, M.~P., {et~al.} 2012, \mnras, 424,
  2180, \dodoi{10.1111/j.1365-2966.2012.21377.x}

\bibitem[{{M{\'e}ndez-Abreu} {et~al.}(2012){M{\'e}ndez-Abreu},
  {S{\'a}nchez-Janssen}, {Aguerri}, {Corsini}, \&
  {Zarattini}}]{Mendez-Abreu2012}
{M{\'e}ndez-Abreu}, J., {S{\'a}nchez-Janssen}, R., {Aguerri}, J.~A.~L.,
  {Corsini}, E.~M., \& {Zarattini}, S. 2012, \apjl, 761, L6,
  \dodoi{10.1088/2041-8205/761/1/L6}

\bibitem[{{Moretti} {et~al.}(2017){Moretti}, {Gullieuszik}, {Poggianti},
  {Paccagnella}, {Couch}, {Vulcani}, {Bettoni}, {Fritz}, {Cava}, {Fasano},
  {D'Onofrio}, \& {Omizzolo}}]{Moretti2017}
{Moretti}, A., {Gullieuszik}, M., {Poggianti}, B., {et~al.} 2017, \aap, 599,
  A81, \dodoi{10.1051/0004-6361/201630030}

\bibitem[{{Nair} \& {Abraham}(2010)}]{Nair2010}
{Nair}, P.~B., \& {Abraham}, R.~G. 2010, \apjs, 186, 427,
  \dodoi{10.1088/0067-0049/186/2/427}

\bibitem[{{Navarro} {et~al.}(1996){Navarro}, {Frenk}, \& {White}}]{Navarro1996}
{Navarro}, J.~F., {Frenk}, C.~S., \& {White}, S. D.~M. 1996, \apj, 462, 563,
  \dodoi{10.1086/177173}

\bibitem[{{Park} \& {Hwang}(2009)}]{park09}
{Park}, C., \& {Hwang}, H.~S. 2009, \apj, 699, 1595,
  \dodoi{10.1088/0004-637X/699/2/1595}

\bibitem[{{Pfenning} {et~al.}(1991){Pfenning}, {Rieger}, \&
  {Schreckenberg}}]{Pfenning1991}
{Pfenning}, T., {Rieger}, H., \& {Schreckenberg}, M. 1991, Journal de Physique
  I, 1, 323, \dodoi{10.1051/jp1:1991134}

\bibitem[{{Poggianti} {et~al.}(2009){Poggianti}, {Fasano}, {Bettoni}, {Cava},
  {Dressler}, {Vanzella}, {Varela}, {Couch}, {D'Onofrio}, {Fritz},
  {Kjaergaard}, {Moles}, \& {Valentinuzzi}}]{poggianti09}
{Poggianti}, B.~M., {Fasano}, G., {Bettoni}, D., {et~al.} 2009, \apjl, 697,
  L137, \dodoi{10.1088/0004-637X/697/2/L137}

\bibitem[{{Reiprich} \& {B{\"o}hringer}(2002)}]{Reiprich02}
{Reiprich}, T.~H., \& {B{\"o}hringer}, H. 2002, \apj, 567, 716,
  \dodoi{10.1086/338753}

\bibitem[{{Rhee} {et~al.}(2017){Rhee}, {Smith}, {Choi}, {Yi}, {Jaff{\'e}},
  {Candlish}, \& {S{\'a}nchez-J{\'a}nssen}}]{Rhee2017}
{Rhee}, J., {Smith}, R., {Choi}, H., {et~al.} 2017, \apj, 843, 128,
  \dodoi{10.3847/1538-4357/aa6d6c}

\bibitem[{{Romano-D{\'\i}az} {et~al.}(2008){Romano-D{\'\i}az}, {Shlosman},
  {Heller}, \& {Hoffman}}]{Romano2008}
{Romano-D{\'\i}az}, E., {Shlosman}, I., {Heller}, C., \& {Hoffman}, Y. 2008,
  \apjl, 687, L13, \dodoi{10.1086/593168}

\bibitem[{{Saha} \& {Elmegreen}(2018)}]{Saha2018}
{Saha}, K., \& {Elmegreen}, B. 2018, \apj, 858, 24,
  \dodoi{10.3847/1538-4357/aabacd}

\bibitem[{{Sarkar} {et~al.}(2021){Sarkar}, {Pandey}, \&
  {Bhattacharjee}}]{Sarkar2021}
{Sarkar}, S., {Pandey}, B., \& {Bhattacharjee}, S. 2021, \mnras, 501, 994,
  \dodoi{10.1093/mnras/staa3665}

\bibitem[{{Sellwood}(2014)}]{Sellwood2014}
{Sellwood}, J.~A. 2014, Reviews of Modern Physics, 86, 1,
  \dodoi{10.1103/RevModPhys.86.1}

\bibitem[{{Sharp} {et~al.}(2006){Sharp}, {Saunders}, {Smith}, {Churilov},
  {Correll}, {Dawson}, {Farrel}, {Frost}, {Haynes}, {Heald}, {Lankshear},
  {Mayfield}, {Waller}, \& {Whittard}}]{Sharp2006}
{Sharp}, R., {Saunders}, W., {Smith}, G., {et~al.} 2006, in Society of
  Photo-Optical Instrumentation Engineers (SPIE) Conference Series, Vol. 6269,
  Society of Photo-Optical Instrumentation Engineers (SPIE) Conference Series,
  ed. I.~S. {McLean} \& M.~{Iye}, 62690G, \dodoi{10.1117/12.671022}

\bibitem[{{Sheth} {et~al.}(2012){Sheth}, {Melbourne}, {Elmegreen}, {Elmegreen},
  {Athanassoula}, {Abraham}, \& {Weiner}}]{Sheth2012}
{Sheth}, K., {Melbourne}, J., {Elmegreen}, D.~M., {et~al.} 2012, \apj, 758,
  136, \dodoi{10.1088/0004-637X/758/2/136}

\bibitem[{{Sheth} {et~al.}(2003){Sheth}, {Regan}, {Scoville}, \&
  {Strubbe}}]{Sheth2003}
{Sheth}, K., {Regan}, M.~W., {Scoville}, N.~Z., \& {Strubbe}, L.~E. 2003,
  \apjl, 592, L13, \dodoi{10.1086/377329}

\bibitem[{{Sheth} {et~al.}(2008){Sheth}, {Spalsbury}, \&
  {Scoville}}]{Sheth2008}
{Sheth}, K., {Spalsbury}, L., \& {Scoville}, N. 2008, in Astronomical Society
  of the Pacific Conference Series, Vol. 390, Pathways Through an Eclectic
  Universe, ed. J.~H. {Knapen}, T.~J. {Mahoney}, \& A.~{Vazdekis}, 426

\bibitem[{{Skibba} {et~al.}(2012){Skibba}, {Masters}, {Nichol}, {Zehavi},
  {Hoyle}, {Edmondson}, {Bamford}, {Cardamone}, {Keel}, {Lintott}, \&
  {Schawinski}}]{Skibba2012}
{Skibba}, R.~A., {Masters}, K.~L., {Nichol}, R.~C., {et~al.} 2012, \mnras, 423,
  1485, \dodoi{10.1111/j.1365-2966.2012.20972.x}

\bibitem[{{Smith} {et~al.}(2004){Smith}, {Saunders}, {Bridges}, {Churilov},
  {Lankshear}, {Dawson}, {Correll}, {Waller}, {Haynes}, \& {Frost}}]{Smith2004}
{Smith}, G.~A., {Saunders}, W., {Bridges}, T., {et~al.} 2004, in Society of
  Photo-Optical Instrumentation Engineers (SPIE) Conference Series, Vol. 5492,
  Ground-based Instrumentation for Astronomy, ed. A.~F.~M. {Moorwood} \&
  M.~{Iye}, 410--420, \dodoi{10.1117/12.551013}

\bibitem[{{Spinoso} {et~al.}(2017){Spinoso}, {Bonoli}, {Dotti}, {Mayer},
  {Madau}, \& {Bellovary}}]{Spinoso17}
{Spinoso}, D., {Bonoli}, S., {Dotti}, M., {et~al.} 2017, \mnras, 465, 3729,
  \dodoi{10.1093/mnras/stw2934}

\bibitem[{{Varela} {et~al.}(2009){Varela}, {D'Onofrio}, {Marmo}, {Fasano},
  {Bettoni}, {Cava}, {Couch}, {Dressler}, {Kj{\ae}rgaard}, {Moles},
  {Pignatelli}, {Poggianti}, \& {Valentinuzzi}}]{Varela2009}
{Varela}, J., {D'Onofrio}, M., {Marmo}, C., {et~al.} 2009, \aap, 497, 667,
  \dodoi{10.1051/0004-6361/200809876}

\bibitem[{{Vera} {et~al.}(2016){Vera}, {Alonso}, \& {Coldwell}}]{Vera2016}
{Vera}, M., {Alonso}, S., \& {Coldwell}, G. 2016, \aap, 595, A63,
  \dodoi{10.1051/0004-6361/201628750}

\bibitem[{{Vulcani} {et~al.}(2018){Vulcani}, {Poggianti}, {Gullieuszik},
  {Moretti}, {Tonnesen}, {Jaff{\'e}}, {Fritz}, {Fasano}, \&
  {Bettoni}}]{vulcani18}
{Vulcani}, B., {Poggianti}, B.~M., {Gullieuszik}, M., {et~al.} 2018, \apjl,
  866, L25, \dodoi{10.3847/2041-8213/aae68b}

\bibitem[{{Wang} {et~al.}(2012){Wang}, {Kauffmann}, {Overzier}, {Tacconi},
  {Kong}, {Saintonge}, {Catinella}, {Schiminovich}, {Moran}, \&
  {Johnson}}]{Wang2012}
{Wang}, J., {Kauffmann}, G., {Overzier}, R., {et~al.} 2012, \mnras, 423, 3486,
  \dodoi{10.1111/j.1365-2966.2012.21147.x}

\bibitem[{{Wozniak} {et~al.}(1995){Wozniak}, {Friedli}, {Martinet}, {Martin},
  \& {Bratschi}}]{Wozniak1995}
{Wozniak}, H., {Friedli}, D., {Martinet}, L., {Martin}, P., \& {Bratschi}, P.
  1995, \aaps, 111, 115

\bibitem[{{Yoon} {et~al.}(2019){Yoon}, {Im}, {Lee}, {Lee}, \& {Lim}}]{Yoon2019}
{Yoon}, Y., {Im}, M., {Lee}, G.-H., {Lee}, S.-K., \& {Lim}, G. 2019, Nature
  Astronomy, 3, 844, \dodoi{10.1038/s41550-019-0799-7}

\bibitem[{{Zana} {et~al.}(2018){Zana}, {Dotti}, {Capelo}, {Mayer}, {Haardt},
  {Shen}, \& {Bonoli}}]{Zana2018}
{Zana}, T., {Dotti}, M., {Capelo}, P.~R., {et~al.} 2018, \mnras, 479, 5214,
  \dodoi{10.1093/mnras/sty1850}

\end{thebibliography}
\bibliographystyle{aasjournal}

\end{document}